# Error Exponents for Variable-length Block Codes with Feedback and Cost Constraints

Barış Nakiboğlu    Robert G. Gallager
Laboratory for Information and Decision Systems
Massachusetts Institute of Technology, Cambridge, MA, 02139
Email: {nakib, gallager }@mit.edu

October 17, 2018

**Abstract**

Variable-length block-coding schemes are investigated for discrete memoryless channels with ideal feedback under cost constraints. Upper and lower bounds are found for the minimum achievable probability of decoding error $P_{e,\min}$ as a function of constraints $R, \mathcal{P}$, and $\overline{\tau}$ on the transmission rate, average cost, and average block length respectively. For given $R$ and $\mathcal{P}$, the lower and upper bounds to the exponent $-(\ln P_{e,\min})/\overline{\tau}$ are asymptotically equal as $\overline{\tau} \to \infty$. The resulting reliability function, $\lim_{\overline{\tau} \to \infty}(-\ln P_{e,\min})/\overline{\tau}$, as a function of $R$ and $\mathcal{P}$, is concave in the pair $(R, \mathcal{P})$ and generalizes the linear reliability function of Burnashev [2] to include cost constraints. The results are generalized to a class of discrete-time memoryless channels with arbitrary alphabets, including additive Gaussian noise channels with amplitude and power constraints.

## 1  Introduction

The information theoretic effect of feedback in communication has been studied since Shannon [16] showed in 1956 that feedback can not increase the capacity $C$ of a discrete memoryless channel (DMC). At about the same time Elias [6] and Chang [15] gave examples showing that feedback could greatly simplify error correction at rates below capacity.

Many of the known results about feedback communication[1] use block coding, i.e., coding in which messages are transmitted sequentially and each message is completely decoded and released to the destination before transmission of the next message begins. Block coding for feedback communication can be further separated into fixed-length and variable-length coding. The codewords in a fixed-length block code all have the same length, but, due to the feedback, the symbols in each codeword can depend on previous channel outputs as well as the choice of transmitted message. For variable-length block codes, the decoding time can also depend dynamically on the previously received symbols. We assume that the feedback is ideal, meaning that it is noiseless, instantaneous, and of unlimited capacity. Thus we can assume that all information available at the receiver is also available at the transmitter, and consequently the transmitter can determine when the receiver decodes each message.

---

[1]Non-block codes, with overlapping messages, sometimes have significant advantages over block coding. Sahai [11] gives an excellent discussion of non-block coding with feedback and compares it to block coding. We restrict ourselves here to block coding.

A widely used quality criterion for fixed-length block codes of a given rate is the error exponent, $\frac{-\ln P_e}{\tau}$, where $P_e$ is the probability of decoding error and $\tau$ is the block length. Dobrushin [5] showed that the sphere-packing exponent (the well known upper bound to the error exponent without feedback) is also an upper bound for fixed-length block coding with feedback on symmetric DMC's. It has been long conjectured that this is also true for non-symmetric DMC's, but the current best upper bound, by Haroutunian [8], is larger than the sphere packing bound in the non-symmetric case.

Variable-length block coding allows the decoding to be delayed under unusually severe noise, thus usually providing a dramatic increase in error exponent. As explained later, the error exponent for a variable-length block code is defined as $\frac{-\ln P_e}{\overline{\tau}}$ where $\overline{\tau}$ is the expected block length.[2] Similarly, the rate[3] is defined as $\frac{\ln M}{\overline{\tau}}$ where $M$ is the size of the message set. The reliability function, $E(R)$, for a class of coding schemes on a given channel is defined as the asymptotic maximum achievable exponent, as $\overline{\tau} \to \infty$, for codes of rate greater than or equal to $R$. Burnashev [2] developed upper and lower bounds to $E(R)$ for variable-length block codes on DMCs with ideal feedback. For DMC's in which all transition probabilities are positive, Burnashev's upper and lower bounds to $E(R)$ are equal. The resulting function $E(R)$ is linear, going from a positive constant at $R = 0$ to 0 at $R = C$.

For DMC's in which not all transition probabilities are positive[4], Burnashev implicitly showed that $P_e = 0$ is asymptotically achievable for all $R < \mathbf{C}$ (i.e., $\mathbf{C}$ is the zero-error capacity of variable-length block codes[5] for such DMC's.)

The main objective of this paper is to generalize Burnashev's results to DMC's subject to a cost criterion. That is, a non-negative cost $\rho_k \geq 0$ is associated with each letter $k$ of the channel input alphabet, $\{0, 1, \ldots, |\mathcal{X}|-1\}$. It is assumed[6] that $\rho_k = 0$ for at least one choice of $k$. The energy in a codeword $X_1, X_2, \ldots, X_\tau$, where $X_n$ is transmitted at time $n$, $1 \leq n \leq \tau$ and $\tau$ is the decoding time, is defined to be $S_\tau = \rho_{X_1} + \cdots + \rho_{X_\tau}$. As explained more fully later, a variable-length block code is defined to satisfy an average cost (power) constraint $\mathcal{P} \geq 0$ if $\mathbf{E}[S_\tau] \leq \mathbf{E}[\tau]\mathcal{P}$. We will find the corresponding reliability function for all $\mathcal{P} \geq 0$. For all DMC's whose transition probabilities are all positive, this reliability function is a concave function of $(R, \mathcal{P})$. If zero transition probabilities exist, then zero error probability can be achieved at all rates below the cost constrained capacity.

Our interest in cost criteria for DMC's is motivated by the desire to separate the effect of cost constraints from that of infinite alphabet size, thus allowing a better understanding of channels such as additive Gaussian noise where these effects are combined. Pinsker [10] considered fixed-length codes for the discrete-time additive white Gaussian noise channel (AWGNC) with feedback. He showed that the sphere-packing exponent upper bounds the error exponent if a fixed upper bound is placed on the energy of each codeword, independent of the noise sample values. Schalkwijk [14] considered the same model but allowed the codeword energy to depend on the noise, subject to an average energy constraint. He developed a simple algorithm for which the error probability decays as a two-fold exponential of the block length (and thus also of the energy). This was an extension of joint work with Kailath [13] where the infinite bandwidth limit of the problem was considered. Kramer [9] later showed that the error probability could be made to decay n-fold exponentially for any $n$ for the infinite bandwidth case.[7]

---

[2] Error exponents can also be defined in various ways for non-block codes, but the interpretation does not correspond perfectly to block coding exponents; see Sahai [11].

[3] Successive messages require independent identically distributed message transmission times, so this rate is the long-term rate at which message bits can be transferred to the receiver.

[4] Trivial outputs that cannot be reached from any input are excluded throughout.

[5] Thus zero-error capacity for variable-length codes with feedback can be strictly larger than zero-error capacity for fixed-length codes with feedback, which in turn can be strictly larger than zero-error capacity without feedback.

[6] The assumption that the minimum cost symbol has cost 0 causes no loss of generality, since otherwise the minimum cost could be trivially subtracted from all symbol costs.

In the following section, we consider a class of variable-length block cods for DMC's with feedback and cost constraints. These generalize the Yamamoto and Itoh [18] codes to allow for cost constraints. We lower bound the achievable error exponent for these codes as a function of constraints $R, \mathcal{P}$, and $\overline{\tau}$ on rate, average cost, and average block length respectively.

In Section 3, we consider all possible variable-length block codes and derive a lower bound on $\overline{\tau}$ as a function of power constraint $\mathcal{P}$, average error probability $P_e$, and message-set size $M$. This is then converted into an upper bound on the error exponent over all codes of given $R, \mathcal{P}$, and $\overline{\tau}$. We show that as $\overline{\tau} \to \infty$, this upper bound coincides with the lower bound of Section 2, thus determining the reliability function in the presence of a cost constraint.

In Section 4 the results are generalized to a broader class of discrete-time memoryless channels that includes AWGNC's with both power and amplitude constraints.

## 2 Achievability: Asymptotically optimum codes

### 2.1 Forward and feedback channel models and cost constraint

The forward channel is assumed to be a DMC of positive capacity with input alphabet $\mathcal{X} = \{0, \ldots, |\mathcal{X}|-1\}$ and output alphabet $\mathcal{Y} = \{0, \ldots, |\mathcal{Y}|-1\}$. The input and output at time $n$ are denoted by $X_n$ and $Y_n$; the $n$-tuples $X_1, \ldots, X_n$ and $Y_1, \ldots, Y_n$ are denoted by $X^n$ and $Y^n$. The feedback channel is *ideal* in the sense that it is discrete and noiseless with an arbitrarily large alphabet size $|\mathcal{Z}|$ (although $|\mathcal{Z}| = |\mathcal{Y}|$ is sufficient). The symbol $Z_n$ sent from the receiver at time $n$ can depend on $Y^n$ and is received without error at the transmitter after $X_n$ and before $X_{n+1}$ is sent. $Z^n$ denotes $Z_1, \ldots, Z_n$.

The forward DMC is defined by the $|\mathcal{X}|$ by $|\mathcal{Y}|$ transition matrix $\{P_{kj}\}$ where, for each time $n$, $P_{kj} = \mathbf{P}\left[Y_n = j \mid X_n = k\right]$. The channel is memoryless in the sense that

$$\mathbf{P}\left[Y_n \mid X^n, Y^{n-1}, Z^{n-1}\right] = \mathbf{P}\left[Y_n \mid X_n\right].$$

For each input letter $k \in \mathcal{X}$, there is a non-negative transmission cost $\rho_k \geq 0$ and at least one $\rho_k$ is zero. The cost $S_\tau$ of transmitting a codeword of length $\tau$ is the sum of the costs of the $\tau$ symbols in the codeword. A cost constraint $\mathcal{P}$ means that $\mathbf{E}\left[S_\tau\right] \leq \mathcal{P}\mathbf{E}\left[\tau\right]$. We usually refer to $\mathcal{P}$ as a *power constraint* and to $S_\tau$ as *energy*. With this definition of power constraint, $\mathcal{P}$ can be seen to upper bound the long-term time-average cost per symbol over a long string of independent successive message transmissions.

### 2.2 Fixed-length block codes with error-or-erasure decoding

We begin with the slightly simpler problem of finding fixed-length block codes for an error-or-erasure decoder, i.e., a decoder which can either decode the message or produce an erasure symbol.

---

[7]In fact, we have recently shown that error probability can be made to decay n-fold exponentially in the finite bandwidth case where $n$ is proportional to the block length.

The objective will be to minimize (or approximately minimize) the error probability while allowing the erasure probability to be much larger than the error probability but still be close to zero. In the following subsection, this error-and-erasure scheme will be converted into a variable-length block-coding scheme by retransmitting the erased messages.

Consider a code of fixed-length $\ell$ containing two phases of length $\ell_1$ and $\ell_2$ respectively. The first phase uses a power constraint $\mathcal{P}_1$ and the second $\mathcal{P}_2$. To meet an overall power constraint $\mathcal{P}$, we require[8] $\ell_1 \mathcal{P}_1 + \ell_2 \mathcal{P}_2 = \ell \mathcal{P}$. Define $\eta$ as $\ell_1/\ell$, so that this power constraint becomes

$$\mathcal{P} = \eta \mathcal{P}_1 + (1-\eta)\mathcal{P}_2$$

Phase 1 consists of a conventional block code without feedback, operating incrementally close to the capacity $\mathbf{C}(\mathcal{P}_1)$ of the channel subject to constraint $\mathcal{P}_1$,

$$\mathbf{C}(\mathcal{P}_1) \triangleq \max_{\phi:\, \sum_k \phi_k \rho_k \leq \mathcal{P}_1} \sum_{k,j} \phi_k P_{kj} \ln \frac{P_{kj}}{\sum_m \phi_m P_{mj}}. \tag{1}$$

Here and throughout, $\phi$ is assumed to be a probability assignment, i.e., $\phi_k \geq 0$ for each $k$ and $\sum_k \phi_k = 1$. The conventional coding theorem for a constrained DMC with fixed block length and no feedback is as follows:[9] for any $\delta_1 > 0$, there is an $\epsilon_1(\delta_1) > 0$ such that, for all large enough $\ell_1$, codes of block length $\ell_1$ exist with $M \geq e^{\ell_1[\mathbf{C}(\mathcal{P}_1) - \delta_1]}$ codewords, each of energy at most $\ell_1 \mathcal{P}_1$ and each with error probability upper bounded by

$$P_{e1} \leq \exp{-\ell_1 \epsilon_1(\delta_1)}.$$

Using such a code in phase 1, the decoder makes a tentative decision at the end of phase 1. The transmitter (knowing the decision via feedback) then sends a binary codeword, $\mathbf{x}_A$ for 'accept' and $\mathbf{x}_R$ for 'reject' in phase 2. Let $P_{RA}$ be the probability that the receiver decodes $\mathbf{x}_A$ given that $\mathbf{x}_R$ is sent. Similarly, $P_{AR}$ is the probability of decoding $\mathbf{x}_R$ given $\mathbf{x}_A$.

If $\mathbf{x}_A$ is decoded, the receiver gives its tentative decision from phase 1 to the user and the overall probability of error $\tilde{P}_e$ satisfies $\tilde{P}_e \leq P_{RA}$. If $\mathbf{x}_R$ is decoded, an erasure is released and the probability of erasure $\tilde{P}_r$ satisfies $\tilde{P}_r \leq P_{AR} + P_{e1}$. Assume for now that the power constraint may be violated by an incrementally small amount. Thus we choose $\mathbf{x}_A$ to satisfy the constraint, and choose $\mathbf{x}_R$ arbitrarily since it is rarely used. We bound $-\ln P_{RA}$ by the divergence between the output distribution conditional on $\mathbf{x}_A$ and the output distribution conditional on $\mathbf{x}_R$.

To be more explicit, define the maximum single-letter divergence for the input letter $k$ as

$$D_k \triangleq \max_m \sum_j P_{kj} \ln \frac{P_{kj}}{P_{mj}}$$

Note that if $P_{mj} = 0$ for some channel transition, then $D_k = \infty$ for each $k$ such that $P_{kj} > 0$. We will see in subsection 2.5 that this leads to error free codes at rates below capacity. In the following subsection, we consider only channels for which $P_{kj} > 0$ for all $k \in \mathcal{X}$ and $j \in \mathcal{Y}$.

---

[8] We could equally well constrain $\mathcal{P}_1, \mathcal{P}_2$ to satisfy $\ell_1 \mathcal{P}_1 + \ell_2 \mathcal{P}_2 \leq \ell \mathcal{P}$, but since these are all inequality constraints, this would simply add an extra degree of freedom into the problem.

[9] See, for example, Theorem 7.3.2 in [7]

### 2.2.1 Error-and-erasure decoding with all $P_{kj} > 0$

Assume that $P_{kj} > 0$ for all $k \in \mathcal{X}$ and $j \in \mathcal{Y}$ and, for each $k \in \mathcal{X}$, let $m_k$ be an input letter $m$ maximizing $\sum_j P_{kj} \ln \frac{P_{kj}}{P_{mj}}$. If $\mathbf{x}_A$ contains $\phi_k \ell_2$ occurrences of letter $k$ and $\mathbf{x}_r$ is chosen to contain the letter $m_k$ whenever $\mathbf{x}_A$ contains $k$, then the following minor variation of Stein's lemma results[10]: for any $\delta_2 > 0$, there is an $\epsilon_2(\delta_2) > 0$ such that

$$P_{RA} \leq \exp\left[\sum_k -\ell_2 \phi_k D_k + \ell_2 \delta_2\right] \tag{2}$$

$$P_{AR} \leq \exp[-\ell_2 \epsilon_2(\delta_2)] \tag{3}$$

From (2), we want to choose $\mathbf{x}_A$ to maximize $\sum_k D_k \phi_k$ subject to the power constraint. Thus, for a power constraint $\mathcal{P}_2$ in phase 2, define $\mathbf{D}(\mathcal{P}_2)$ as

$$\mathbf{D}(\mathcal{P}_2) \triangleq \max_{\phi:\ \sum_k \phi_k \rho_k \leq \mathcal{P}_2} \sum_k D_k \phi_k. \tag{4}$$

The function $\mathbf{D}(\mathcal{P})$ in (4) is the maximum of a linear function of $\phi$ over linear constraints. As illustrated in Figure 1, $\mathbf{D}(\mathcal{P})$ is piecewise linear, non-decreasing, and concave in its domain of definition, $\mathcal{P} \geq 0$. Choosing the phase 2 codewords $\mathbf{x}_A$ and $\mathbf{x}_R$ according to this maximization, (2)

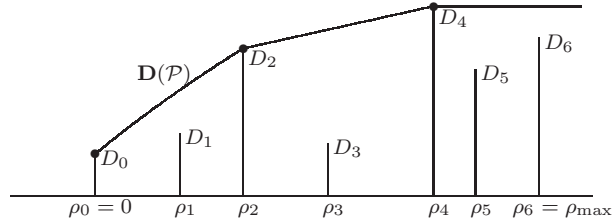

Figure 1: The function $\mathbf{D}(\mathcal{P})$ for a channel satisfying $P_{kj} > 0$ for all $k \in \mathcal{X}$ and $j \in \mathcal{Y}$. The maximum single-letter divergences $D_k$ are also shown. For convenience, the inputs are ordered in terms of cost. For any given $\mathcal{P}$, $\mathbf{D}(\mathcal{P})$ can be achieved with at most 2 positive $\phi_k$.

becomes

$$P_{RA} \leq \exp[-\ell_2 \mathbf{D}(\mathcal{P}_2) + \ell_2 \delta_2] \tag{5}$$

The power constraint $\mathcal{P}_2$ is then satisfied by $\mathbf{x}_A$. The power in $\mathbf{x}_R$ (whose probability of usage vanishes exponentially with $\ell_2$) can be upper bounded by $\rho_{\max}$. The preceding results are summarized in the following lemma.

**Lemma 1** *For all $\mathcal{P}_1 \geq 0$, $\mathcal{P}_2 \geq 0$, $0 < \eta < 1$, $\delta_1 > 0$, $\delta_2 > 0$, and all sufficiently large $\ell$, there is an error-and-erasure code with $M \geq \exp\{\eta\ell[\mathbf{C}(\mathcal{P}_1) - \delta_1]\}$ such that, for each message, the probability of error $\tilde{P}_e$, the probability of erasure $\tilde{P}_r$, and the expected energy $\mathbf{E}[\mathcal{S}]$ satisfy*

$$\tilde{P}_e \leq \exp\{-(1-\eta)\ell\left[\mathbf{D}(\mathcal{P}_2) - \delta_2\right]\} \tag{6}$$

$$\tilde{P}_r \leq e^{-\eta\ell\epsilon_1(\delta_1)} + e^{-(1-\eta)\ell\,\epsilon_2(\delta_2)}, \tag{7}$$

$$\mathbf{E}[\mathcal{S}] \leq \ell[\eta\mathcal{P}_1 + (1-\eta)\mathcal{P}_2 + \rho_{\max}e^{-\eta\ell\epsilon_1(\delta_1)}] \tag{8}$$

---

[10] This can be derived, for example, by starting with Theorem 5 in [4] and specializing to the case of asymptotically small $s$.

## 2.3 Variable-length block codes; all $P_{kj} > 0$

The above error-or-erasure code can form the basis of a variable-length block code with ideal feedback. As in Yamamoto and Itoh [18], the transmitter observes each erasure via the feedback and repeats the original message until it is accepted. For simplicity, we assume that when a message is repeated, the receiver ignores the previous received symbols and uses the same decoding algorithm as before. Since an error occurs independently after each repetition of the fixed length codeword, the overall error probability satisfies

$$P_e \leq \frac{1}{1 - \tilde{P}_r} \exp\left\{-(1-\eta)\ell[\mathbf{D}(\mathcal{P}_2) - \delta_2]\right\}.$$

The duration $\tau$ of a block is $\ell$ times the number of error-or-erasure tries until acceptance, so $\mathbf{E}[\tau] = \ell/(1 - \tilde{P}_r)$. The coefficient $1/(1 - \tilde{P}_r)$ goes to 1 with increasing $\ell$ and thus can be absorbed into the arbitrary term $\delta_2$ for sufficiently large $\ell$. Similarly $\ell$ can be replaced with $\overline{\tau} = \mathbf{E}[\tau]$, yielding $P_e \leq \exp\left\{-\overline{\tau}(1-\eta)[\mathbf{D}(\mathcal{P}_2) - \delta_2]\right\}$.

Similarly the expected energy $\mathbf{E}[\mathcal{S}_\tau]$ over the entire transmission satisfies $\mathbf{E}[\mathcal{S}_\tau] \leq \mathbf{E}[\mathcal{S}]/(1 - \tilde{P}_r)$. Finally, using (8), the average power for each codeword is

$$\frac{\mathbf{E}[\mathcal{S}_\tau]}{\mathbf{E}[\tau]} \leq \eta \mathcal{P}_1 + (1-\eta)\mathcal{P}_2 + \rho_{\max} e^{-\eta \ell \epsilon_1(\delta_1)}$$

The following lemma summarizes these results

**Lemma 2** *Assume ideal feedback for a DMC with all $P_{kj} > 0$. For all $\eta \in (0,1), \mathcal{P}_1 \geq 0, \mathcal{P}_2 \geq 0, \delta > 0$, and sufficiently large $\overline{\tau}$, there is a variable-length block code with $M \geq \exp\{\overline{\tau}[\eta\mathbf{C}(\mathcal{P}_1) - \delta]\}$ messages, each using average power at most $\eta\mathcal{P}_1 + (1-\eta)\mathcal{P}_2 + \delta$, and each with error probability*

$$P_e \leq \exp\{-\overline{\tau}[(1-\eta)\mathbf{D}(\mathcal{P}_2) - \delta]\} \tag{9}$$

## 2.4 Optimization of the bound; all $P_{kj} > 0$

Lemma 2 can be interpreted as providing a nominal rate of transmission, $R = \eta\mathbf{C}(\mathcal{P}_1)$, a nominal power constraint, $\mathcal{P} = \eta\mathcal{P}_1 + (1-\eta)\mathcal{P}_2$, and a nominal exponent of error probability, $(1-\eta)\mathbf{D}(\mathcal{P}_2)$. We have demonstrated the existence of variable-length block codes for which the actual average rate, power, and exponent approach these values arbitrarily closely as $\overline{\tau}$ becomes large.

For any given $\mathcal{P}$ and $R$ satisfying[11] $0 < R < \mathbf{C}(\mathcal{P})$, we now maximize the exponent $(1-\eta)\mathbf{D}(\mathcal{P}_2)$ over $0 < \eta < 1$, $\mathcal{P}_1 \geq 0$, and $\mathcal{P}_2 \geq 0$, subject to the constraints $R = \eta\mathbf{C}(\mathcal{P}_1)$ and $\eta\mathcal{P}_1 + (1-\eta)\mathcal{P}_2 = \mathcal{P}$. Our strategy, for a given $\eta$, will be to use $R = \eta\mathbf{C}(\mathcal{P}_1)$ to specify $\mathcal{P}_1$ and then use $\eta\mathcal{P}_1 + (1-\eta)\mathcal{P}_2 = \mathcal{P}$ to specify $\mathcal{P}_2$, which is constrained by $\mathcal{P}_2 \geq 0$. Satisfying these constraints will put some constraints on $\eta$, and the exponent will then be a function of $R, \mathcal{P}$, and $\eta$. The maximization then reduces to a maximization over the single constrained parameter $\eta$.

The constraints on $\eta$ and the ensuing maximization depend on the properties of the capacity function $\mathbf{C}(\mathcal{P})$ illustrated in Figure 2. As can be seen from (1) and visualized in Figure 2, the function $\mathbf{C}(\mathcal{P})$ is non-negative, concave, continuous, and non-decreasing for all $\mathcal{P} \geq 0$. It is strictly increasing for $0 \leq \mathcal{P} \leq \mathcal{P}^*$ where $\mathcal{P}^*$ is the smallest $\mathcal{P}$ for which $\mathbf{C}(\mathcal{P}) = \mathbf{C}^*$, the unconstrained channel capacity. This suggests that $\mathcal{P}_1$ is determined from $\eta$. The following two lemmas, which are proven in the appendix, make this precise.

---

[11]The interesting special case where $R = \mathbf{C}(\mathcal{P})$ is discussed in the following subsection.

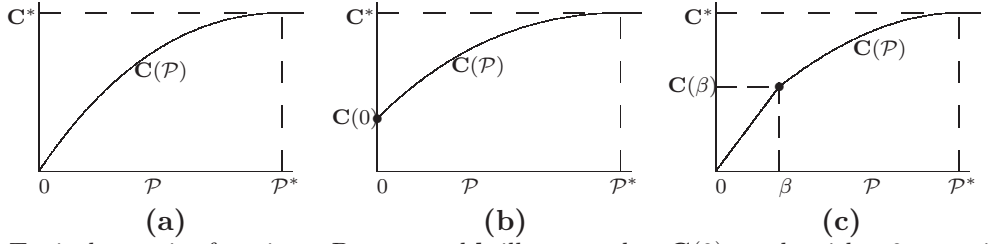

Figure 2: Typical capacity functions. Parts **a** and **b** illustrate that $\mathbf{C}(0)$ can be either 0 or positive. Part **c** illustrates an important special case where $\mathbf{C}(x)$ is linear from 0 to $\beta > 0$ where $\beta$ is defined as the largest $x$ for which $\mathbf{C}(y)/y = \mathbf{C}(x)/x$ for all $y \in (0, x]$.

**Lemma 3** *For any $R, \mathcal{P}$ such that $0 < R < \mathbf{C}(\mathcal{P})$, the equation $\eta \mathbf{C}(\mathcal{P}/\eta) = R$ has a unique solution for $\eta \in (0, 1)$; that solution, say $\eta^*_{R,\mathcal{P}}$, satisfies $\eta^*_{R,\mathcal{P}} \in [\frac{R}{C^*}, \frac{R}{C(0)}]$.*

The solution, $\eta^*(R, \mathcal{P})$ can also be expressed explicitly as $\eta^*_{R,\mathcal{P}} = \frac{\mathcal{P}}{\Gamma^{-1}(R/\mathcal{P})}$, where $\Gamma^{-1}(\cdot)$ is the inverse of the function $\Gamma(x) = \mathbf{C}(x)/x$ taken over the domain $x \geq \beta$, where $\beta$ is the largest $x$ for which $\Gamma(x) = \Gamma(0)$.

For any $0 < R < \mathbf{C}(\mathcal{P}$, define

$$\mathcal{I}_{R,\mathcal{P}} \triangleq \left[\eta^*_{R,\mathcal{P}}, \ \min\left(1, \frac{R}{\mathbf{C}(0)}\right)\right]. \tag{10}$$

**Lemma 4** *For any $R, \mathcal{P}$ such that $0 < R < \mathbf{C}(\mathcal{P})$ and for any $\eta \in \mathcal{I}_{R,\mathcal{P}}$, the following properties hold:*

- *There is a unique $\mathcal{P}_1 \in [0, \mathcal{P}^*]$ such that $R = \eta \mathbf{C}(P_1)$.*

- *The corresponding $\mathcal{P}_2$, defined by $\mathcal{P} = \eta \mathcal{P}_1 + (1-\eta)\mathcal{P}_2$, is nonnegative.*

*There is no $\mathcal{P}_1, \mathcal{P}_2 \geq 0$ such that $R = \eta \mathbf{C}(P_1)$ and $\mathcal{P} = \eta \mathcal{P}_1 + (1-\eta)\mathcal{P}_2$ for $\eta \notin \mathcal{I}_{R,\mathcal{P}}$.*

Thus for any $(R, \mathcal{P})$ pair such that $0 < R < \mathbf{C}(\mathcal{P})$, and for any $\eta \in \mathcal{I}_{R,\mathcal{P}}$, the nominal exponent is

$$E(R, \mathcal{P}, \eta) = (1 - \eta)\mathbf{D}\left(\frac{\mathcal{P} - \eta \mathbf{C}^{-1}(R/\eta)}{1 - \eta}\right) \tag{11}$$

The following lemma, proved in the appendix, shows that $E(R, \mathcal{P}, \eta)$ is concave.

**Lemma 5** *The set of points $(R, \mathcal{P}, \eta)$ such that $0 < R < \mathbf{C}(\mathcal{P})$ and $\eta \in \mathcal{I}_{R,\mathcal{P}}$ is convex. The function $E(R, \mathcal{P}, \eta)$ is concave over this domain.*

We next maximize the exponent $E(R, \mathcal{P}, \eta)$ over $\eta \in \mathcal{I}_{R,\mathcal{P}}$,

$$E(R, \mathcal{P}) \triangleq \sup_{\eta \in \mathcal{I}_{R,\mathcal{P}}} (1 - \eta)\mathbf{D}\left(\frac{\mathcal{P} - \eta \mathbf{C}^{-1}(R/\eta)}{1 - \eta}\right) \tag{12}$$

This is simply a concave maximization over an interval. The resulting function, $E(R, \mathcal{P})$ is then also concave as a function of $(R, \mathcal{P})$, and thus also as a function of $R$ for any given $\mathcal{P}$. This is illustrated in Figure 3. It can be shown that $E(R, \mathcal{P})$ is strictly decreasing in $R$ from $\mathbf{D}(\mathcal{P})$ at $R = 0$.

One can extend the definition of $E(R, \mathcal{P})$ to $R = \mathbf{C}(\mathcal{P})$, for any $\mathcal{P}$ as

$$E(\mathbf{C}(\mathcal{P}), \mathcal{P}) \triangleq \lim_{\delta \to 0^+} E(\mathbf{C}(\mathcal{P}) - \delta, \mathcal{P}) \tag{13}$$

The following theorem results from using $E(R, \mathcal{P})$ in Lemma 2.

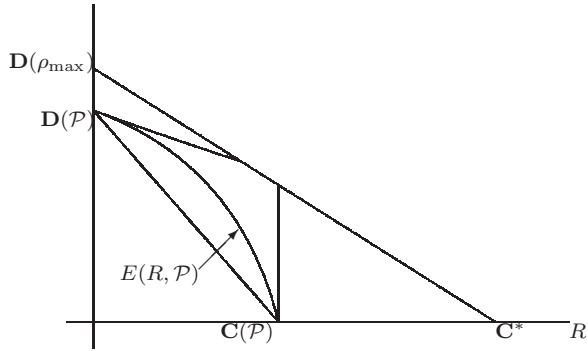

Figure 3: A typical $E(R, \mathcal{P})$ curve. The figure illustrates that $E(R, \mathcal{P})$, as a function of $R$ for fixed $\mathcal{P}$, is concave, decreasing, and bounded.

**Theorem 1** *Assume ideal feedback for a DMC with all $P_{kj} > 0$. Then for all $0 < R \leq \mathbf{C}(\mathcal{P})$, all positive $\delta$, and all sufficiently large integer $\ell$, there is a variable-length block code of expected length $\overline{\tau}$, $\ell \leq \overline{\tau} < \ell + 1$ with $M \geq \exp[\overline{\tau}(R - \delta)]$ messages such that for each message, $\theta \in \mathcal{M}$ the probability of error $P_e(\theta)$ and the expected energy $\mathbf{E}[\mathcal{S}_\tau(\theta)]$ satisfy*

$$P_e(\theta) \leq \exp\{-\overline{\tau}[E(R, \mathcal{P}) - \delta]\} \tag{14}$$

$$\mathbf{E}[\mathcal{S}_\tau(\theta)] \leq \left(\mathcal{P} + \rho_{\max} e^{-\overline{\tau}\epsilon(\delta)}\right) \overline{\tau}, \tag{15}$$

*where $\epsilon(\delta) > 0$ for each $\delta > 0$. Furthermore, the probability that the codeword length exceeds $\ell$ is at most $\delta$.*

Theorem 1 shows that the exponent $E(R, \mathcal{P})$ can be asymptotically achieved by this particular class of variable-length block codes. The converse in the next section will show that no variable-length block code can do better asymptotically, i.e., that $E(R, \mathcal{P})$ is the reliability function for constrained variable-length block codes.

Theorem 1 also shows that these codes are almost fixed-length block codes, deviating from fixed length only with arbitrarily small probability. It is also possible to analyze the queuing delay for this class of codes. Note that if the source bits arrive equally spaced in time, then, even for a fixed-length block code, bits are delayed waiting for the next block and additionally delayed waiting for the block to be received and decoded. The additional delay, for variable-length block codes, is the queuing delay of waiting blocks while earlier blocks are retransmitted. At any $R < \mathbf{C}$, the probability of retransmission decreases exponentially (albeit with a small exponent) with $\overline{\tau}$, so it is not surprising that the expected additional delay due to retransmissions goes to 0 with increasing $\overline{\tau}$. We have shown that this indeed happens.

For $0 < R < \mathbf{C}(\mathcal{P})$ and $\mathcal{P} > 0$, (15) can be simplified by absorbing the term $\rho_{\max} e^{-\bar{\tau}\epsilon(\delta)}$ into the $\delta$ of (14). This cannot be done for $\mathcal{P} = 0$ since the constraint $\mathbf{E}\left[\mathcal{S}_\tau(\theta)\right] \leq \mathcal{P}\bar{\tau}$ for all $\theta$ reduces to the unconstrained case where only zero-cost inputs are used. In (15), on the other hand, we are using a reject messages of positive power with asymptotically vanishing probability, with increasing $\bar{\tau}$.

The requirement of ideal feedback can be relaxed to that of a noiseless feedback link of capacity $\mathbf{C}_{\text{fb}} \geq \mathbf{C}$ and finite delay $\mathbf{T}$ by using a modification of the error-and erasure scheme first suggested by Şimşek and Sahai[12] for unconstrained channels. For phase 1, the message is divided into equal length sub-messages which are separately encoded at a rate close to capacity and sent one after the other. A temporary decision about each sub-message is made at the receiver and sent reliably to the transmitter with a delay equal to $\mathbf{T}$ plus the sub-message transmission time. In phase 2, the entire message is rejected if any sub-message was in error and otherwise it is accepted. A single bit of feedback is required for phase 2, and it can be shown that the various delays become amortized over the entire message transmission as $M \to \infty$.

### 2.4.1 Channels for which $E(R, \mathcal{P}) > 0$ for $R = \mathbf{C}(\mathcal{P})$

In certain cases $E(\mathbf{C}(\mathcal{P}), \mathcal{P})$ as defined in equation (13) is strictly positive.[12] We start with a simple example of this phenomenon and then delineate the cases where it is possible.

**Example 1: BSC with extra free symbol:** Consider a binary symmetric channel (BSC) in which each input symbol has unit cost. There is an additional cost-free symbol that is completely noisy. That is, the transition probabilities and costs are as follows:

$$P_{kj} = \begin{bmatrix} 1/2 & 1/2 \\ \alpha & 1-\alpha \\ 1-\alpha & \alpha \end{bmatrix} \qquad \rho_k = \begin{bmatrix} 0 \\ 1 \\ 1 \end{bmatrix} \tag{16}$$

where $0 < \alpha < 1/2$. Letting $\mathbf{C}_{\text{BSC}} = \ln 2 - \mathfrak{h}(x)$ where $\mathfrak{h}(x) = -(1-\alpha)\ln(1-\alpha) - \alpha \ln \alpha$ is the binary entropy, it can be seen that[13]

$$\mathbf{C}(\mathcal{P}) = \begin{cases} \mathcal{P}\mathbf{C}_{\text{BSC}} & \text{for} \quad 0 \leq \mathcal{P} \leq 1 \\ \mathbf{C}_{\text{BSC}} & \text{for} \quad \mathcal{P} > 1. \end{cases}$$

Assume a power constraint $\mathcal{P} = 1/2$ and rate $R = \mathbf{C}(\mathcal{P}) = \frac{1}{2}\mathbf{C}_{\text{BSC}}$. One half unit of energy is required in phase 1 and the maximum interval for phase 2 is provided by choosing $\mathcal{P}_1 = 1$ and $\eta = 0.5$. Thus we transmit at the unconstrained capacity during phase 1 and transmit at zero power for phase 2. In phase 2, the zero symbol is used for $\mathbf{x}_A$ and either of the unit cost symbols is used for $\mathbf{x}_R$. This yields an exponent $E(R, \mathcal{P}) = \frac{1}{2}D_0 = \frac{1}{4}\ln(\frac{1}{4\alpha(1-\alpha)})$, which is clearly positive. What is happening is that zero nominal power (i.e., zero power except for the rare transmission of reject symbols) provides a positive exponent.

Now consider $E(R, \mathcal{P})$ for this channel for any $R < \mathbf{C}(\mathcal{P})$ and $\mathcal{P} \leq 1$. It can be seen that $\eta^*_{R,\mathcal{P}} = R/\mathbf{C}_{\text{BSC}}$. It can also be seen, as above, that this value of $\eta$, corresponding to $\mathcal{P}_1 = 1$, maximizes $E(R, \mathcal{P}, \eta)$. Thus

$$E(R, \mathcal{P}) = (1-\eta)\mathbf{D}(\frac{\mathcal{P} - \eta}{1 - \eta}) \qquad \text{where } \eta = \frac{R}{\mathbf{C}_{\text{BSC}}}$$

---

[12]This result contains the epsilons and deltas of Theorem 1, and thus does not assert reliable transmission 'at capacity', but the existence of a positive exponent at $R = \mathbf{C}(\mathcal{P})$ is still very surprising.

[13]The linearity of the capacity function here follows from the fact that the output probabilities in (1) are the same for the free symbol and the equiprobable use of the BSC symbols.

From (4), $\mathbf{D}(x) = D_0 + (D_1 - D_0)x$ for $0 \leq x \leq 1$, so

$$E(R, \mathcal{P}) = (1 - \eta)D_0 + (\mathcal{P} - \eta)(D_1 - D_0) \qquad \text{where } \eta = \frac{R}{\mathbf{C}_{\text{BSC}}}$$

This is illustrated in Figure 4. Consider the limit of $E(R, \mathcal{P})$ as $R$ approaches $\mathbf{C}(\mathcal{P})$ from below and thus $\eta$ approaches $\mathbf{C}(\mathcal{P})/\mathbf{C}_{\text{BSC}} = \mathcal{P}$. This limit is $(1 - \mathcal{P})D_0$ and is asymptotically achievable at $R = \mathbf{C}(\mathcal{P})$ by transmitting with power 1 in phase 1 and then transmitting with a nominal power equal to 0 during phase 2.

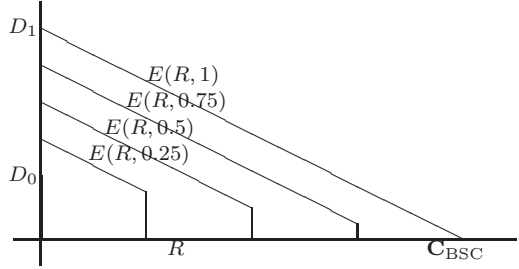

Figure 4: $E(R, \mathcal{P})$ for a BSC with a zero cost noise symbol. For $\mathcal{P} < 1$, the exponent decreases linearly to a positive value at capacity.

More generally, any DMC for which $\beta > 0$ has the property that if $\mathcal{P} < \beta$, then $E(R, \mathcal{P})$ is positive at $R = \mathbf{C}(\mathcal{P})$ and affine in $R$ for $0 < R < \mathbf{C}(\mathcal{P})$.

### 2.4.2 Alternative approaches to finding $E(R, \mathcal{P})$

The reliability function $E(R, \mathcal{P})$ is expressed in (12) as an optimization over $E(R, \mathcal{P}, \eta)$ and as such involves calculating $\mathbf{C}(\mathcal{P}_1)$ and $\mathbf{D}(\mathcal{P}_2)$ as subproblems. An alternative that might be more convenient numerically is to express $E(R, \mathcal{P})$ directly as a concave optimization over the input probabilities in phase 1 and 2 subject to the constraints corresponding to a given $R$ and $\mathcal{P}$.

Another alternative, which is more interesting conceptually, is to investigate how the phase 1 and phase 2 powers must be related. Consider the equivalent problem of finding the minimal power $\mathcal{P}$ required for a given rate $R$ and exponent $E$. We will derive a necessary condition for $\mathcal{P}_1 > 0$, $\mathcal{P}_2 > 0$, and $0 < \eta < 1$ to achieve this minimum power. First consider the special case in which $\mathbf{C}(\mathcal{P})$ is continuously differentiable for $\mathcal{P} > 0$ and let $A_1 = \eta \mathcal{P}_1$ be the phase 1 power amortized over both phases. The partial derivative of $A_1$ with respect to $\eta$ for a given $R = \eta \mathbf{C}(A_1/\eta)$ is then

$$\left.\frac{\partial A_1}{\partial \eta}\right|_R = -\frac{\partial [\eta \mathbf{C}(A_1/\eta)]/\partial \eta}{\partial [\eta \mathbf{C}(A_1/\eta)]/\partial A_1} = \mathcal{P}_1 - \frac{\mathbf{C}(\mathcal{P}_1)}{\mathbf{C}'(\mathcal{P}_1)} \tag{17}$$

Geometrically, this is the horizontal axis intercept of the tangent to $\mathbf{C}(\cdot)$ at $\mathcal{P}_1$.

In the general case, $\mathbf{C}(\mathcal{P}_1)$ can have slope discontinuities at particular values of $\mathcal{P}_1$; because of these discontinuities the left and right derivatives and the corresponding tangents and intercepts becomes different from each other (see Figure 5).

In the same way, let $A_2 = (1 - \eta)\mathcal{P}_2$. Then, holding the exponent $E$ fixed,

$$\left.\frac{\partial A_2}{\partial \eta}\right|_E = -\frac{\partial [(1 - \eta)\mathbf{D}(A_2/[(1 - \eta)]/\partial \eta}{\partial [[(1 - \eta)\mathbf{D}(A_2/[(1 - \eta))]/\partial A_2} = -\mathcal{P}_2 + \frac{\mathbf{D}(\mathcal{P}_2)}{\mathbf{D}'(\mathcal{P}_2)} \tag{18}$$

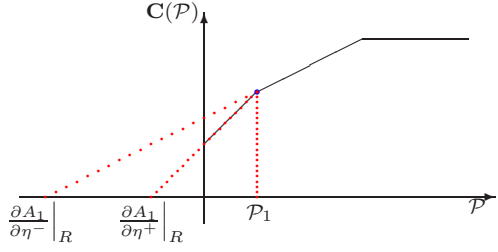

Figure 5: $\frac{\partial A_1}{\partial \eta^-}|_R$ is the derivative corresponding to negative change in $\eta$ (positive change in $\mathcal{P}_1$) and $\frac{\partial A_1}{\partial \eta^+}|_R$ is the derivative corresponding to positive change in $\eta$ (negative change in $\mathcal{P}_1$).

This is the negative of the horizontal axis intercept of the tangent to $\mathbf{D}(\cdot)$ at $\mathcal{P}_2$. At points of slope discontinuity in $\mathbf{D}(\cdot)$, this must be replaced with

$$\left.\frac{\partial A_2}{\partial \eta^-}\right|_E = -\mathcal{P}_2 + \frac{\mathbf{D}(\mathcal{P}_2)}{d[\mathbf{D}(\mathcal{P}_2)]/d[\mathcal{P}_2^-]}; \qquad \left.\frac{\partial A_2}{\partial \eta^+}\right|_E = -\mathcal{P}_2 + \frac{\mathbf{D}(\mathcal{P}_2)}{d[\mathbf{D}(\mathcal{P}_2)]/d[\mathcal{P}_2^+]}$$

Finally, the overall power constraint is $\mathcal{P} = A_1 + A_2$, so

$$\left.\frac{\partial \mathcal{P}}{\partial \eta^+}\right|_{R,E} = \left.\frac{\partial A_1}{\partial \eta^+}\right|_R + \left.\frac{\partial A_2}{\partial \eta^+}\right|_E; \qquad \left.\frac{\partial \mathcal{P}}{\partial \eta^-}\right|_{R,E} = \left.\frac{\partial A_1}{\partial \eta^-}\right|_R + \left.\frac{\partial A_2}{\partial \eta^-}\right|_E$$

For $\mathcal{P}_1, \mathcal{P}_2$, and $\eta$ to minimize $\mathcal{P}$ for fixed $R, E$, it is necessary that $\frac{\partial \mathcal{P}}{\partial \eta^+}|_{R,E} \geq 0$ (i.e., that an incremental increase in $\eta$ does not reduce $\mathcal{P}$) and that $\frac{\partial \mathcal{P}}{\partial \eta^-}|_{R,E} \leq 0$ (i.e., that an incremental decrease in $\eta$ does not reduce $\mathcal{P}$). Geometrically, what this says is that the horizontal intercept of the tangent to $\mathbf{C}(\cdot)$ at $\mathcal{P}_1$, which in general is the interval $[\frac{\partial A_1}{\partial \eta^-}|_R, \frac{\partial A_1}{\partial \eta^+}|_R]$, must overlap with the horizontal intercept of the tangent to $\mathbf{D}(\cdot)$ at $\mathcal{P}_2$, i.e., with the interval $[-\frac{\partial A_2}{\partial \eta^+}|_E, -\frac{\partial A_2}{\partial \eta^-}|_E]$. Note that these intervals reduce to single points in the absence of slope discontinuities in $\mathbf{C}(\mathcal{P}_1)$ or $\mathbf{D}(\mathcal{P}_2)$.

It is surprising that these conditions do not involve $\eta$. The following example shows how these conditions can be used.

**Example 2: Combined 4-input symmetric channel, BSC, and free symbol:** Consider the following DMC with seven input letters and four output letters;

$$P_{kj} = \begin{bmatrix} 1/4 & 1/4 & 1/4 & 1/4 \\ \delta & \delta & 1/2-\delta & 1/2-\delta \\ 1/2-\delta & 1/2-\delta & \delta & \delta \\ 1-3\epsilon & \epsilon & \epsilon & \epsilon \\ \epsilon & 1-3\epsilon & \epsilon & \epsilon \\ \epsilon & \epsilon & 1-3\epsilon & \epsilon \\ \epsilon & \epsilon & \epsilon & 1-3\epsilon \end{bmatrix} \qquad \rho_k = \begin{bmatrix} 0 \\ 1 \\ 1 \\ 4 \\ 4 \\ 4 \\ 4 \end{bmatrix}$$

where $\epsilon = 1/75$, $\delta = 1/100$.

$\mathbf{C}(\mathcal{P})$ is piecewise linear for the same reason as in the previous example; $\mathbf{C}(\mathcal{P})$ and $\mathbf{D}(\mathcal{P})$ are given in Figure 6.

The above necessary conditions on $\mathcal{P}_1$ and $\mathcal{P}_2$ imply

$$\begin{aligned} \mathcal{P}_1 = 1 &\Leftrightarrow 1 > \mathcal{P}_2 > 0 \\ 4 > \mathcal{P}_1 > 1 &\Leftrightarrow \mathcal{P}_2 = 1 \\ \mathcal{P}_1 = 4 &\Leftrightarrow \mathcal{P}_2 > 1 \end{aligned} \qquad (19)$$

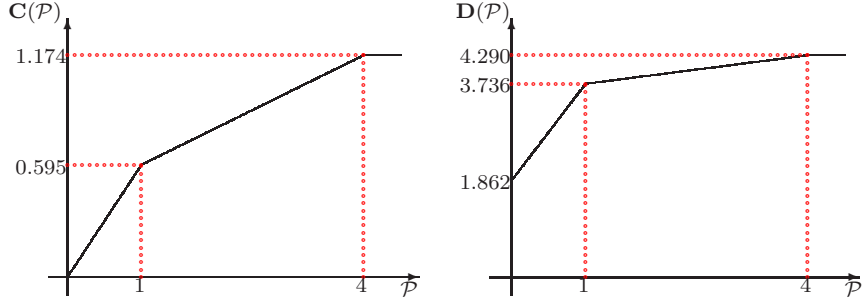

Figure 6: Capacity and Divergence functions, to different scales

Using these conditions and the set constraint, we can calculate $E(R,\mathcal{P})$ for any given $\mathcal{P}$; the solutions for $\mathcal{P}=1$ and $\mathcal{P}=5$ are given in Figure 7.

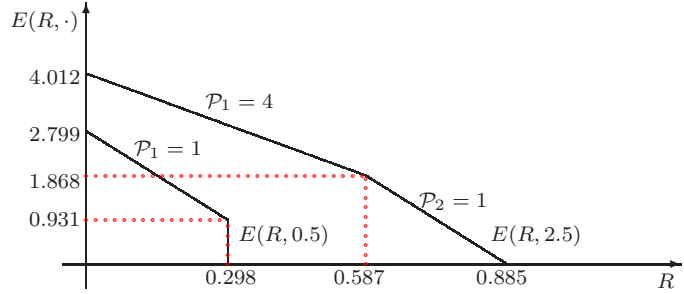

Figure 7: Reliability function for $\mathcal{P}=0.5$ and $\mathcal{P}=2.5$; each straight line segment is characterized by either constant $\mathcal{P}_1$ or constant $\mathcal{P}_2$ according to (19).

## 2.5 Zero-error capacity; Channels with at least one $P_{kj}=0$

The form of $E(R,\mathcal{P})$ relies heavily on the assumption that $P_{kj}>0$ for all $k,j$. To see why, assume $P_{mj}=0$ for some $m,j$. Since all outputs are assumed to be reachable, $P_{kj}$ must be strictly positive for some $k$ and that same $j$. In this case $D_k=\infty$. Suppose that the 'accept' codeword of section 2 uses all $k$'s, the 'reject' message all $m$'s, and that the receiver decodes 'accept' only if it receives one or more $j$'s. In this case, no errors can ever occur for the corresponding variable-length block code.

Asymptotically, phase 2 can occupy a negligible portion of the block, say $\ln\ell$ of $\ell$ symbols. Then for any $\delta>0$, and all large enough block lengths $\ell$, an error-and-erasure code exists with $M\geq e^{(\ell-\ln\ell)(\mathbf{C}(\mathcal{P})-\delta)}$, $\tilde{P}_e=0$, and $\tilde{P}_r\leq e^{-(\ell-\ln\ell)\epsilon(\delta)}+\ell^{\ln(1-\mathsf{q})}$ and expected energy $\mathbf{E}\left[\mathcal{S}_\ell\right]\leq \ell\mathcal{P}+\rho_{\max}\ln\ell$. After a little analysis the following theorem results:

**Theorem 2** *Assume ideal feedback for a DMC with at least one $P_{kj}=0$. Then for all $0<R\leq \mathbf{C}(\mathcal{P})$, all positive $\delta$, and all sufficiently large $\overline{\tau}$, there is a variable-length block code satisfying*

$$M\geq e^{\overline{\tau}(R-\delta)} \qquad P_e=0 \qquad \mathbf{E}\left[\mathcal{S}_\tau\right]\leq \mathcal{P}\overline{\tau}+2\rho_{\max}\ln\overline{\tau}$$

# 3 The converse: relating $\overline{\tau}$ and $P_e$

We have established an upper bound on $P_e$ for given rate $R$, power $\mathcal{P}$, and expected block length $\overline{\tau}$ by developing and analyzing a particular class of variable-length block codes. Here we develop a lower bound to $P_e$ which, for large enough $\overline{\tau}$, is valid for all variable length block codes. The lower bound uses the idea of a two phase analysis, but, as will be seen, this does not restrict the encoding or decoding. We start by finding a lower bound on the expected time $\mathbf{E}[\tau_1]$ spent in the first phase and a related lower bound on the expected time $\mathbf{E}[\tau - \tau_1]$ spent in the second phase.

The analysis is a simplification and generalization of Burnashev [2] and is based on the evolution at each time $n$ of the conditional message entropy, conditioned on the observations at the receiver. The first phase is the interval until this conditional entropy drops from $\ln M$ to some fixed intermediate value, taken here to be 1. The second phase is the interval until this conditional entropy further drops to meet the constraint on error probability; Fano's inequality is used to link the conditional entropy to the error probability. In the first phase we create a stochastic sequence related to the decrease in conditional entropy at each instant $n$, and in the second phase we create a stochastic sequence related to the decrease in the logarithm of the conditional entropy.

Establishing this lower bound to $P_e$ is more involved than the upper bound to $P_e$, since the lower bound must apply to *all* variable-length block codes. We start with a more precise definition of variable-length block codes. After that we bound the expected change of conditional entropy and its logarithm, first in one time unit and second between two stopping times. Then these are used to lower bound the probability of error. The resulting upper bound on the reliability function agrees with the lower bound in section 2.

## 3.1 Mathematical preliminaries and Fano's inequality

In a variable-length block code, the transmitter is assumed to initially receive one of $M$ equiprobable messages from the set $\mathcal{M} = \{1, \ldots, M\}$. It transmits successive channel symbols about that message, say message $\theta$, until the receiver makes a decision and releases the decoded message to the user. The time of this decision is a random variable denoted by $\tau$. We assume throughout that $\mathbf{E}[\tau] = \overline{\tau} < \infty$, since otherwise any desired lower bound to $\overline{\tau}$ is obviously satisfied.

Given noiseless feedback, we can restrict our attention to encoding algorithms in which each input symbol $X_n$ is a deterministic function of message and feedback.[14]

$$X_n = X_n(\theta, Z^{n-1}) \qquad \forall Z^{n-1}, \forall \theta. \tag{20}$$

The entire observation of the receiver up to time $n$, including $Y^n$ and any additional random choices, can be summarized by the $\sigma$-field $\mathcal{F}_n$ generated by these random variables. The nested sequence of $\mathcal{F}_n$'s is called a filtration $\mathcal{F}$.

At each time $n$, depending on the realization $\mathfrak{f}_n$ of $\sigma$-field $\mathcal{F}_n$, the receiver has an a posteriori probability $p_i(\mathfrak{f}_n)$ for each $i$ in $\mathcal{M}$. The corresponding conditional entropy of the message, given $\mathcal{F}_n$, is a random variable $\mathcal{H}_{\mathcal{F}_n}$, measurable in $\mathcal{F}_n$. Its sample value for any realization $\mathfrak{f}_n \in \mathcal{F}_n$, is given by:

$$\mathcal{H}_{\mathfrak{f}_n} = H(\theta \mid \mathcal{F}_n = \mathfrak{f}_n) = -\sum_{i=1}^{M} p_i(\mathfrak{f}_n) \ln p_i(\mathfrak{f}_n).$$

---

[14] This allows the receiver to feed back not only the channel outputs but also some random choices. Random choices at the transmitter provide no added generality since those choices (for all possible $\theta$) could be made earlier at the receiver with no loss of performance.

A decoding algorithm includes a decision rule about continuing or stopping the communication, depending on the observations up to that time, i.e., a Markov stopping time with respect to the filtration $\mathcal{F}$. The message is also decoded at this stopping time. In order to define the various random variables at all times $n \geq 1$, rather than only times up to the stopping time, we will assume that $X_n(\theta, Z^{n-1})$ is equal to some given zero-cost symbol for all $n > \tau$ and all $\theta$. Thus $S_n = S_\tau$ for all $n > \tau$. Thus if a variable-length block code (henceforth simply called a code) satisfies a cost constraint $\mathbf{E}[S_\tau] < \mathcal{P}\overline{\tau}$, then $\mathbf{E}[S_\tau] < \infty$ and $\mathbf{E}[S_n] < \infty$ for all $n$.

Fano's inequality can be applied for each $\mathfrak{f}_\tau$ to upper bound the conditional entropy $\mathcal{H}_{\mathfrak{f}_\tau}$ in terms of the error probability of the decoding at $\mathfrak{f}_\tau$. Taking the expectation[15] of these inequalities over $\mathfrak{f}_\tau \in \mathcal{F}_\tau$, and using the concavity of the binary entropy, $\mathfrak{h}(x)$, the expected value of $\mathbf{E}[\mathcal{H}_{\mathcal{F}_\tau}]$ can be upper bounded at the decoding time by

$$\begin{aligned} \mathbf{E}[\mathcal{H}_{\mathcal{F}_\tau}] &\leq \mathfrak{h}(P_e) + P_e \ln(M-1) \\ &\leq P_e(\ln M - \ln P_e + 1). \end{aligned} \quad (21)$$

This suggests that the conditional entropy is usually very small at the decoding time, motivating a focus on how fast the logarithm of the entropy changes in the second phase of the analysis below.

### 3.2 Bounds on the change of conditional entropy

For any DMC, any code, and any $\mathcal{P} \geq 0$, define the stochastic sequence $\{\mathbf{V}_n; n \geq 0\}$ as

$$\mathbf{V}_n \triangleq \mathcal{H}_{\mathcal{F}_n} + n\mathbf{C}(\mathcal{P}) + \gamma_\mathbf{C}^\mathcal{P}(\mathbf{E}[\mathcal{S}_n | \mathcal{F}_n] - n\mathcal{P}) \quad (22)$$

where $\gamma_\mathbf{C}^\mathcal{P} \geq 0$ is the Lagrange multiplier for the cost constraint in the maximization of $\mathbf{C}(\mathcal{P})$ over input probabilities in (1). The random variable $\mathbf{V}_n$ will be used to bound the entropy and energy changes in phase 1.

**Lemma 6** *For any DMC, any code, and any $\mathcal{P} \geq 0$, the sequence $\{\mathbf{V}_n; n \geq 0\}$ is a submartingale, i.e.,*

$$\mathbf{E}[|\mathbf{V}_n|] < \infty \quad \text{and} \quad \mathbf{E}[\mathbf{V}_{n+1} | \mathcal{F}_n] \geq \mathbf{V}_n \quad \text{for all} \quad n \geq 0.$$

This lemma applies to all codes, whether or not they have a cost constraint equal to the $\mathcal{P}$ in the definition of $\mathbf{V}_n$. Proofs of Lemmas 6 to 10 are given in the appendix.

The following two lemmas develop another submartingale based on the log entropy.

**Lemma 7** *For any DMC with all $P_{kj} > 0$, any code, and any $n \geq 0$,*

$$\mathbf{E}\left[\ln \mathcal{H}_{\mathcal{F}_n} - \ln \mathcal{H}_{\mathcal{F}_{n+1}} \middle| \mathcal{F}_n\right] \leq \mathbf{E}\left[\mathbf{D}_{X_{n+1}} \middle| \mathcal{F}_n\right] \quad (23)$$

Another stochastic sequence, $\{\mathbf{W}_n; n \geq 0\}$ is now defined[16] that combines the changes in log entropy and cost.

$$\mathbf{W}_n \triangleq \ln \mathcal{H}_{\mathcal{F}_n} + n\mathbf{D}(\mathcal{P}) + \gamma_\mathbf{D}^\mathcal{P}(\mathbf{E}[\mathcal{S}_n | \mathcal{F}_n] - n\mathcal{P}), \quad (24)$$

where $\gamma_\mathbf{D}^\mathcal{P} \geq 0$ is the Lagrange multiplier for the cost constraint in the maximization of $\mathbf{D}(\mathcal{P})$ over input probabilities in (4).

---

[15]The facts that $|\mathcal{H}_{\mathcal{F}_\tau}| \leq \ln M$ and $|P_e| \leq 1$, combined with Lebesgue's dominated convergence theorem, allow us to interchange the limit and expectation here.

[16] Note that $\mathbf{W}_n$ can be $-\infty$ for DMC's in which $P_{kj} = 0$ for one or more transitions, but $\mathbf{W}_n$ will be used only for DMC's in which all $P_{kj} > 0$.

**Lemma 8** *For any DMC with all $P_{kj} > 0$, for any code, and for any $\mathcal{P}$, $0 \leq \mathcal{P} < \infty$, the sequence $\{\mathbf{W}_n; n \geq 0\}$ is a submartingale, i.e.,*

$$\mathbf{E}\left[|\mathbf{W}_n|\right] < \infty \quad and \quad \mathbf{E}\left[\mathbf{W}_{n+1} | \mathcal{F}_n\right] \geq \mathbf{W}_n \quad \text{for all} \quad n \geq 0.$$

### 3.3 Measuring Time with Submartingales

The following lemmas are used to lower bound the expected stopping times for phases 1 and 2 in terms of $\{\mathbf{V}_n; n \geq 0\}$ and $\{\mathbf{W}_n; n \geq 0\}$

**Lemma 9** *For any DMC and any code, if a stopping time $\tau_1$ satisfies*

$$\mathbf{E}\left[\tau_1\right] < \infty \quad and \quad \mathbf{E}\left[\mathcal{S}_{\tau_1}\right] \leq \mathcal{P} \mathbf{E}\left[\tau_1\right]$$

*then*

$$\mathbf{C}(\mathcal{P})\mathbf{E}\left[\tau_1\right] \geq \mathbf{E}\left[\mathcal{H}_{\mathcal{F}_0} - \mathcal{H}_{\mathcal{F}_{\tau_1}}\right] \tag{25}$$

**Lemma 10** *For any DMC with all $P_{kj} > 0$ and any code, if a pair of stopping times $(\tau_1 \leq \tau_2)$ satisfies,*

$$\mathbf{E}\left[\tau_2\right] < \infty \quad and \quad \mathbf{E}\left[\mathcal{S}_{\tau_2} - \mathcal{S}_{\tau_1}\right] \leq \mathcal{P} \mathbf{E}\left[\tau_2 - \tau_1\right]$$

*then*

$$\mathbf{D}(\mathcal{P})\mathbf{E}\left[\tau_2 - \tau_1\right] \geq \mathbf{E}\left[\ln \mathcal{H}_{\mathcal{F}_{\tau_1}} - \ln \mathcal{H}_{\mathcal{F}_{\tau_2}}\right] \tag{26}$$

The bounds asserted by these lemmas are tight in the sense that when they are used with the stopping times to be specified later, they will show that $E(R, \mathcal{P})$ in (12) is an upper bound on the reliability function.

### 3.4 Lower bounding $\overline{\tau}$ for DMC's with all $P_{kj} > 0$

We now derive lower bounds on the expected decoding time for any variable-length block code with $M$ equiprobable messages, subject to a given cost constraint $\mathcal{P}$ and a required probability of error $P_e$. The first result is simply an explicit statement of the well-known impossibility of transmitting reliably at rates above $\mathbf{C}(\mathcal{P})$.

**Theorem 3** *For any DMC, any code with $M \geq 2$ equiprobable messages, any $\mathcal{P} \geq 0$, and any required error probability $P_e \geq 0$ and cost constraint $\mathcal{P}$, the expected decoding time satisfies*

$$\mathbf{E}\left[\tau\right] \geq \frac{\ln M - P_e(\ln M - \ln P_e + 1)}{C(\mathcal{P})} \tag{27}$$

**Proof:** From (21), $\mathbf{E}\left[\mathcal{H}_{\mathfrak{f}_\tau}\right] \leq P_e(\ln M - \ln P_e + 1)$. Thus, since $\mathcal{H}_{\mathfrak{f}_0} = \ln M$, (27) results from (25) with $\tau_1 = \tau$. **QED**

This result is valid both for the case where all $P_{kj} > 0$ and the zero-error case where some $P_{kj} = 0$. In the zero-error case, we already know that $P_e = 0$ is asymptotically achievable for $R < \mathbf{C}(\mathcal{P})$, so our remaining task is to show that $E(R, \mathcal{P})$ is an upper bound as well as a lower bound to the reliability function in the case where $R < \mathbf{C}(\mathcal{P})$ and all $P_{kj} > 0$.

### 3.4.1 Lower bounding $\bar{\tau}$ for DMC's with all $P_{kj} > 0$

The main issue in this lower bound is finding an intermediate Markov stopping time $\tau_1$ which will divide the message transmission interval into two disjoint phases[17] such that the duration of each can be lower bounded by Lemmas 9 and 10 respectively. Consider the stopping time $t_1 = \min\{n \mid \mathcal{H}_{\mathcal{F}_n} \leq 1\}$ in filtration $\mathcal{F}$. This does not quite work as an intermediate stopping time, since a variable-length code could in principle occasionally decode before $\mathcal{H}_{\mathcal{F}_n} \leq 1$. Instead we use $\tau_1 = \min(\tau, t_1)$ to define the end of the first phase. This is also a Markov stopping time, and $0 \leq \tau_1 \leq \tau$, so this is a well defined intermediate time for all codes.

We now apply Lemma 9 to $\tau_1$. Let $\mathbf{E}[\mathcal{S}_{\tau_1}]$ be the expected energy used by any given code in this first phase and let $\mathcal{P}_1 = \frac{\mathbf{E}[\mathcal{S}_{\tau_1}]}{\mathbf{E}[\tau_1]}$. Then (25) becomes

$$\mathbf{E}\left[\mathcal{H}_{\mathcal{F}_0} - \mathcal{H}_{\mathcal{F}_{\tau_1}}\right] \leq \mathbf{C}(\mathcal{P}_1)\mathbf{E}[\tau_1]. \tag{28}$$

We first find an upper bound to $\mathbf{E}\left[\mathcal{H}_{\mathcal{F}_{\tau_1}}\right]$. By definition of $t_1$, $\mathcal{H}_{\mathcal{F}_{t_1}} \leq 1$, but $\mathcal{H}_{\mathcal{F}_{\tau_1}}$ might be greater than 1 if $\mathcal{H}_{\mathcal{F}_\tau} > 1$. Thus we can upper bound $\mathbf{E}\left[\mathcal{H}_{\mathcal{F}_{\tau_1}}\right]$ by

$$\begin{aligned}
\mathbf{E}\left[\mathcal{H}_{\mathcal{F}_{\tau_1}}\right] &\leq 1 + \mathbf{P}\left[\mathcal{H}_{\mathcal{F}_\tau} > 1\right] \mathbf{E}\left[\mathcal{H}_{\mathcal{F}_{\tau_1}} \mid \mathcal{H}_{\mathcal{F}_\tau} > 1\right] \\
&\leq 1 + \mathbf{P}\left[\mathcal{H}_{\mathcal{F}_\tau} > 1\right] \ln M & (29) \\
&\leq 1 + \mathbf{E}\left[\mathcal{H}_{\mathcal{F}_\tau}\right] \ln M & (30) \\
&\leq 1 + P_e(\ln M - \ln P_e + 1) \ln M, & (31)
\end{aligned}$$

where in (29) we upper bounded $\mathbf{E}\left[\mathcal{H}_{\mathcal{F}_{\tau_1}} \mid \mathcal{H}_{\mathcal{F}_\tau} > 1\right]$ by $\ln M$, the maximum entropy for any ensemble of $M$ elements. We used the Markov inequality in (30) and then (21) in (31).

Since the messages are a priori equiprobable, $\mathcal{H}_{\mathcal{F}_0} = \ln M$, so substituting this and (31) into (28),

$$\mathbf{E}[\tau_1] \geq \frac{\ln M \left[1 - P_e(\ln M - \ln P_e + 1) - \frac{1}{\ln M}\right]}{\mathbf{C}(\mathcal{P}_1)}. \tag{32}$$

As shown later, the term in brackets essentially approaches 1 as $P_e \to 0$ and thus $\mathbf{E}[\tau_1]$ is approximately lower bounded by $(\ln M)/\mathbf{E}[\mathbf{C}(\mathcal{P}_1)]$.

Next we find the expected time $\mathbf{E}[\tau - \tau_1]$ spent in phase 2. Here we use (26) from Lemma 10, with the initial time $\tau_i$ chosen to be $\tau_1$ and the final time $\tau_f$ chosen to be $\tau$. Let $\mathbf{E}[\mathcal{S}_\tau - \mathcal{S}_{\tau_1}]$ be the expected energy used by the given code in this second phase and let $\mathcal{P}_2 = \frac{\mathbf{E}[\mathcal{S}_\tau - \mathcal{S}_{\tau_1}]}{\mathbf{E}[\tau - \tau_1]}$. Then

$$\mathbf{E}\left[\ln \mathcal{H}_{\mathcal{F}_{\tau_1}} - \ln \mathcal{H}_{\mathcal{F}_\tau}\right] \leq \mathbf{D}(\mathcal{P}_2)\mathbf{E}[\tau - \tau_1] \tag{33}$$

We lower-bound $\mathbf{E}\left[\ln \mathcal{H}_{\mathcal{F}_{\tau_1}} - \ln \mathcal{H}_{\mathcal{F}_\tau}\right]$ by upper-bounding $\mathbf{E}\left[\ln \mathcal{H}_{\mathcal{F}_\tau}\right]$ and lower-bounding $\mathbf{E}\left[\ln \mathcal{H}_{\mathcal{F}_{\tau_1}}\right]$. By Jensen's inequality, $\mathbf{E}\left[\ln \mathcal{H}_{\mathcal{F}_\tau}\right] \leq \ln \mathbf{E}\left[\mathcal{H}_{\mathcal{F}_\tau}\right]$, so from (21),

$$\mathbf{E}\left[\ln \mathcal{H}_{\mathcal{F}_\tau}\right] \leq \ln[P_e(\ln M - \ln P_e + 1)]. \tag{34}$$

To lower-bound $\mathbf{E}\left[\ln \mathcal{H}_{\mathcal{F}_{\tau_1}}\right]$, we use the following lemma,

---

[17]There is a nice intuitive relation between these two phases used in the converse and the two phases used in the variable-length block codes of Section 2.3, since in each case the first phase deals with a large sea of messages and the second deals essentially with a binary hypothesis. When an error-and-erasure codeword is repeated, however, phase 1 as defined here could end during any one of those repetitions.

**Lemma 11** *For any DMC with all $P_{kj} > 0$, any code, and any $n \geq 0$,*

$$\ln \mathcal{H}_{\mathcal{F}_n} - \ln \mathcal{H}_{\mathcal{F}_{n+1}} \leq \max_{k,m,j} \ln \frac{P_{kj}}{P_{mj}} \triangleq \mathbf{F} \tag{35}$$

Since $\mathcal{H}_{\mathcal{F}_{\tau_1-1}} > 1$, i.e., $\ln \mathcal{H}_{\mathcal{F}_{\tau_1-1}} > 0$, the lemma implies that $\ln \mathcal{H}_{\mathcal{F}_{\tau_1}} \geq -\mathbf{F}$. Substituting this and (34) into (33),

$$\mathbf{E}\left[\tau - \tau_1\right] \geq \frac{-\ln P_e - \mathbf{F} - \ln[\ln M - \ln P_e + 1]}{\mathbf{D}(\mathcal{P}_2)} \tag{36}$$

As shown later, the numerator is essentially $(-\ln P_e)$ in the limit of small $P_e$. Now we can find a lower bound on $(-\ln P_e)/\bar{\tau}$, for codes of rate $(\ln M)/\bar{\tau}$, using the above result.

**Theorem 4** *Assume a DMC with all $P_{kj} > 0$. Let $\mathcal{P} \geq 0$, $0 \leq R \leq \mathbf{C}(\mathcal{P})$, and $\delta > 0$ be arbitrary. Then, for all sufficiently large $\bar{\tau}$, all variable-length block codes with*

- *expected energy $\mathbf{E}\left[\mathcal{S}_\tau\right] \leq \mathcal{P}\bar{\tau} + \delta$*
- *$M \geq \exp[\bar{\tau}(R + \delta)]$ equiprobable messages*

*must satisfy*

$$P_e \geq \exp\{-\bar{\tau}[E(R,\mathcal{P}) + \delta].\} \tag{37}$$

We now give an intuitive justification of the theorem.; a proof is given in the appendix. Leaving out the 'negligible' terms, (32) and (36) are

$$\bar{\tau}_1 \geq \frac{\ln(M)}{\mathbf{C}(\mathcal{P}_1)} \quad ; \quad \bar{\tau} - \bar{\tau}_1 \geq \frac{-\ln(P_e)}{\mathbf{D}(\mathcal{P}_2)}.$$

Defining $\eta = \frac{\bar{\tau}_1}{\bar{\tau}}$ for the given code and rearranging terms,

$$\frac{\ln(M)}{\bar{\tau}} \leq \eta \mathbf{C}(\mathcal{P}_1) \quad ; \quad \frac{-\ln(P_e)}{\bar{\tau}} \leq (1-\eta)\mathbf{D}(\mathcal{P}_2). \tag{38}$$

For given $\mathcal{P}_1, \mathcal{P}_2$, and $\eta$, (38) is the converse of Lemma 2. The exponent $E(R,\mathcal{P})$ is the result of optimizing over these parameters for given $R, \mathcal{P}$. The proof in the appendix treats the neglected quantities and this optimization carefully.

## 4 Extension to Other Memoryless Channels

The channel model of Sections 2 and 3 assumes finite input and output alphabets, but, as will be seen, the analysis is more general, and with some assumptions, continues to hold with minor changes such as replacing sums and max's with integrals and sup's. A later paper by Burnashev[18] [3], extends his results for the DMC to more general memoryless channels, and the results to follow generalize this to channels with cost constraints. In this section, we specify a family of channels with cost constraints in which Theorems 1 and 4 are valid, i.e., for which upper and lower bounds to the reliability function are equal to $E(R,\mathcal{P})$.

---

[18]The authors are grateful to the reviewers of [1] and to Peter Berlin for pointing out this reference.

## 4.1 Assumptions about the channel model

The channel input and output alphabets can be countable or uncountably infinite and will be denoted by $\mathcal{X}$ and $\mathcal{Y}$ respectively. Each element $x \in \mathcal{X}$ has an associated cost $\rho_x$ and as before we assume that the infimum of these costs is equal to zero.

Each input $x \in \mathcal{X}$ will have an associated probability measure $\vartheta_x$ governing the output conditional on input $x$; this replaces the transition matrix $\{P_{kj}\}$ for the DMC case. We will assume that there exists a probability measure $\nu$, with respect to which all $\vartheta_x$ are absolutely continuous.

$$\forall x \in \mathcal{X} \qquad \nu >> \vartheta_x$$

Indeed without this $\nu$ one can hardly begin to analyze such a memoryless channel. For each $x \in \mathcal{X}$, let $\psi_x$ be the Radon-Nikodym derivative[19] of $\vartheta_x$ with respect to $\nu$.

$$\psi_x = \frac{\vartheta_x}{\nu} \tag{39}$$

Our previous definitions can be extended by replacing sums with integrals, max's with sup's etc.

$$\mathbf{C}(\mathcal{P}) \triangleq \sup_{\mu: \int_{\mathcal{X}} \rho_x d\mu \leq \mathcal{P}} \int_{\mathcal{X}} d\mu \int_{\mathcal{Y}} \psi_x \ln \frac{\psi_x}{\psi_\mu} d\nu \tag{40}$$

where $\psi_\mu = \int_{\mathcal{X}} \psi_x d\mu$ and $\mu$ is the unconditional probability measure on $\mathcal{Y}$. Similarly,

$$\mathbf{D}_x \triangleq \sup_{\alpha \in \mathcal{X}} \int_{\mathcal{Y}} \psi_x \ln \frac{\psi_x}{\psi_\alpha} d\nu \tag{41}$$

$$\mathbf{D}(\mathcal{P}) \triangleq \sup_{\mu: \int_{\mathcal{X}} \rho_x d\mu \leq \mathcal{P}} \int_{\mathcal{X}} \mathbf{D}_x d\mu \tag{42}$$

The following assumption will ensure that $\mathbf{D}(\mathcal{P})$ and $\mathbf{C}(\mathcal{P})$ are finite for all $\mathcal{P} \geq 0$.

**Assumption 1** *The discrete time memoryless channel satisfies the following:*[20]

- $\forall x \in \mathcal{X}$, $\mathbf{D}_x < \infty$ and $\rho_x < \infty$
- $\forall \mathcal{P} > 0$, $\Lambda(\mathcal{P}) \triangleq \sup_{x: \rho_x \leq \mathcal{P}} \mathbf{D}_x < \infty$
- $\limsup_{\mathcal{P} \to \infty} \frac{\Lambda(\mathcal{P})}{\mathcal{P}} < \infty$

Using the above assumption in place of the DMC assumption, It is straightforward to verify that the proof of Theorem 1 still holds.[21]

Proceeding on to the converse, it can be seen that the proofs of Lemmas 6 to 10 all hold under Assumption 1. In verifying these proofs, however, one must assume that all codes have finite expected energy; this is tacitly assumed in Theorem 4 since we are assuming throughout that $\overline{\tau} < \infty$. Lemma 11 does not hold in all cases, and in particular does not even hold for the amplitude limited AWGNC. The following additional assumption will hold in many cases where Lemma 11 does not hold and will enable Theorem 4 to be proven.

---

[19] If $\mathcal{X}$ and $\mathcal{Y}$ are each the set of real numbers, and if probability densities exist, then $\psi_x(y)$ can be taken as the probability density of $y$ conditional on $x$.

[20] This assumption is clearly satisfied for every DMC with all $P_{kj} > 0$. If some $P_{kj}$ is zero, then $\mathbf{D}(\mathcal{P})$ is infinite for all $\mathcal{P} > 0$ and the assumption is not satisfied.

[21] Some additional $\delta$'s are required because of the supremums in the definitions, but they can be made negligible by increasing $\ell$.

**Assumption 2** *The discrete-time memoryless channel has an associated function $\xi(\cdot)$ such that:*

- *For any coding and any n*

$$\mathbf{E}\left[\left[\ln \mathcal{H}_{\mathcal{F}_n} - \ln \mathcal{H}_{\mathcal{F}_{n+1}}\right]_{(a)} \big| \mathcal{F}_n\right] \leq \xi(a)\left(1 + \mathbf{E}\left[(\mathcal{S}_{n+1} - \mathcal{S}_n)| \mathcal{F}_n\right]\right) \quad (43)$$

- $\lim_{a \to \infty} \xi(a) = 0$

where $[\cdot]_{(a)} = \cdot \mathbb{I}_{\{\cdot \geq a\}}$.

Theorem 4 is proved for stationary memoryless channels satisfying Assumptions (1) and (2) in the appendix.

## 4.2 Discussion of extended channel models

It is natural at this point to ask what kinds of channels satisfy Assumptions 1 and 2. A partial answer comes from considering the class of channels without cost constraints considered by Burnashev in [3]. He shows that any channel satisfying the following conditions has an error exponent given by $E = (1 - R/\mathbf{C})\mathbf{D}$.

- $\mathbf{D} = \sup_{\alpha,\beta \in \mathcal{X}} \int \psi_\alpha \ln\left(\frac{\psi_\alpha}{\psi_\beta}\right) d\nu < \infty$

- $\phi(a) = \sup_{\alpha,\beta \in \mathcal{X}} \int \psi_\alpha \ln\left(\frac{\psi_\alpha}{\psi_\beta}\right) \mathbb{I}_{\left\{\ln\left(\frac{\psi_\alpha}{\psi_\beta}\right) > a\right\}} d\nu < \infty$, and $\lim_{a \to \infty} \phi(a) = 0$

- At least one of the following is satisfied

  - The channel is an additive noise channel whose input alphabet is a closed interval on the real line and whose noise has a unimodal density.

  - $\exists K > 0$ such that $\sup_{\alpha,\beta \in \mathcal{X}} \int \psi_\alpha \left(\ln \frac{\psi_\alpha}{\psi_\beta}\right)^{(1+K)} d\nu < \infty$

His first assumption is that $\sup_{x \in \mathcal{X}} \mathbf{D}_x < \infty$. This implies that the channel satisfies our Assumption 1 for all non-negative finite cost assignments. He shows that the other assumptions imply that a function $\xi(a)$, $a \geq 0$ exists such that $\lim_{a \to \infty} \xi(a) = 0$ and such that for all codes,

$$\mathbf{E}\left[\left[\ln \mathcal{H}_{\mathcal{F}_n} - \ln \mathcal{H}_{\mathcal{F}_{n+1}}\right]_{(a)} \big| \mathcal{F}_n\right] \leq \xi(a) \quad (44)$$

This implies that the channel satisfies our Assumption 2 for all non-negative finite cost assignments. Thus his assumptions imply our assumptions for all cost assignments and thus imply that, for all cost assignments, the corresponding $E(R, \mathcal{P})$ exists and is the reliability function.

We next give an example of a channel that does not satisfy the Burnashev conditions and does not have a finite reliability function without a cost constraint, but does satisfy our conditions and has a cost constrained reliability function $E(R, \mathcal{P})$.

Let $\mathcal{X}$ and $\mathcal{Y}$ be the set of non-negative integers and assume the cost function $\rho_x = x^2$, i.e., the cost of each input letter is equal to the square of the value of the correspond real number. Let the transition probability $P_{xy}$ be

$$P_{xy} = \frac{1}{3}\left(\frac{1}{2^y} + \delta[x - y]\right)$$

where $\delta[\cdot]$ is 1 when its argument is 0, and 0 elsewhere.

This channel can be proved to satisfy Assumptions 1 and 2, and thus its error exponent is given by $E(R, \mathcal{P})$. On the other hand it does not satisfy the necessary conditions of [3]. Furthermore, the reliability function is unbounded for any rate below capacity if there is no cost constraint.

## 5 Conclusions

Theorems 1 and 4 specify the reliability function for the class of variable-length block codes for DMC's with cost constraints, all $P_{kj} > 0$, and ideal[22] feedback. The results are extended to a more general class of discrete-time memoryless channels satisfying Assumptions 1 and 2 of Section 4. AWGNC's with amplitude and power constraints provide examples satisfying these assumptions. Theorem 2 shows that zero error probability is achievable at all rates up to the cost constrained capacity and moreover is achievable by a very simple scheme.

The rate and the error exponent are specified in terms of the expected block length. By looking at a long sequence of successive message transmissions, it is evident from the law of large numbers that the rate corresponds to the average number of message bits transmitted per unit time. In the same way, the cost constraint is satisfied as an average over both time and channel behavior. The theorems then say essentially that the probability of error $P_{e,\min}$ for the best variable-length block code of given $R, \mathcal{P}$, and $\overline{\tau}$ satisfies $\frac{-\ln P_{e,\min}}{\overline{\tau}} \to E(R, \mathcal{P})$ as $\overline{\tau} \to \infty$.

Mathematically these theorems are quite similar to the conventional non-feedback block coding results except for the following differences: first, the reliability function is known for all rates rather than rates sufficiently close to capacity; second the reliability function is concave (and sometimes positive at capacity); and third the reliability function is given in terms of expected rather than actual block length. The first two differences have been discussed in detail in the previous sections.

In order to understand the role of the expected block length on the exponent, look at the coding scheme used for achievability. $\overline{\tau}$ is close to the fixed block length of the underlying error-and-erasure code and it is this code that determines $E(R, \mathcal{P})$. In other words, the variable-length feature is essential for the small error probability, but $\tau$ is constant with high probability.

One might think that a variable-length block code has many system disadvantages over a fixed-length code, but this is not really true (except for time-sensitive systems) since variable-length protocols are almost invariably used at all higher layers. As discussed in Section 2, it can be shown that the expected additional queuing delay introduced by those codes can be made to approach 0 with increasing $\overline{\tau}$.

## A Proofs

**Proof of Lemma 3:**

As can be seen from Figure 2, $\frac{\mathbf{C}(x)}{x}$ is the slope of the straight line between the points $(0,0)$ and $(x, \mathbf{C}(x))$. It is constant for $0 \leq x \leq \beta$ and is continuous and strictly decreasing for $x \geq \beta$. Here $\beta \geq 0$ is the largest $x$ for which $\mathbf{C}(y)/y = \mathbf{C}(x)/x$ for all $y \in (0, x)$; $\beta = 0$ for cases (a) and (b) of the figure. Letting $x = \mathcal{P}/\eta$ shows that $\eta \mathbf{C}(\mathcal{P}/\eta)$, as a function of $\eta > 0$, is continuous and strictly increasing for $\eta \leq \mathcal{P}/\beta$ and constant for $\eta \geq \mathcal{P}/\beta$. It is equal to 0 at $\eta = 0$, and to $\mathbf{C}(\mathcal{P})$ at $\eta = 1$.

---

[22]Indeed, as argued previously noiseless feedback of rate $\mathbf{C}$ or higher with bounded delay is enough

Since $0 < R < \mathbf{C}(\mathcal{P})$, there is a unique $\eta^*_{R,\mathcal{P}} \in (0, 1)$ satisfying $\eta^*_{R,\mathcal{P}}\mathbf{C}(\mathcal{P}/\eta^*_{R,\mathcal{P}}) = R$ (and thus also satisfying $\eta^*_{R,\mathcal{P}} < \mathcal{P}/\beta$). Since $\mathbf{C}(0) \leq \mathbf{C}(x) \leq \mathbf{C}^*$ for all $x \geq 0$, we have $\mathbf{C}(0) \leq \mathbf{C}(\mathcal{P}/\eta^*_{R,\mathcal{P}}) \leq \mathbf{C}^*$. Thus $\mathbf{C}(0) \leq R/\eta^*_{R,\mathcal{P}} \leq \mathbf{C}^*$ and $\eta^*_{R,\mathcal{P}}$ satisfies $\frac{R}{\mathbf{C}^*} \leq \eta^*_{R,\mathcal{P}} \leq \frac{R}{\mathbf{C}(0)}$. **QED**

**Proof of Lemma 4:**

Note that $\mathbf{C}(x)$ is continuous and strictly increasing from $\mathbf{C}(0)$ at $x = 0$ to $\mathbf{C}^*$ at $x = \mathcal{P}^*$ (see Figure 2). Thus, for each $\eta$ in the interval $[\frac{R}{\mathbf{C}^*}, \frac{R}{\mathbf{C}(0)}]$, $R = \eta\mathbf{C}(\mathcal{P}_1)$ has a unique solution[23] for $\mathcal{P}_1 \leq \mathcal{P}^*$. Since $\eta^*_{R,\mathcal{P}} \in [\frac{R}{\mathbf{C}^*}, \frac{R}{\mathbf{C}(0)}]$, it follows that $\mathcal{I}_{R,\mathcal{P}} \subseteq [\frac{R}{\mathbf{C}^*}, \frac{R}{\mathbf{C}(0)}]$, so $R = \eta\mathbf{C}(\mathcal{P}/\eta)$ has a unique solution for all $\eta \in \mathcal{I}_{R,\mathcal{P}}$, establishing the first itemized property for $\mathcal{I}_{R,\mathcal{P}}$.

Next we must show that that $\eta\mathcal{P}_1 \leq \mathcal{P}$ for the $\mathcal{P}_1 \leq \mathcal{P}^*$ satisfying $R = \eta\mathbf{C}(\mathcal{P}_1)$. Since $\eta \geq \eta^*_{R,\mathcal{P}}$ by assumption, the monotonicity of $\eta\mathbf{C}(\mathcal{P}/\eta)$ implies that $\eta\mathbf{C}(P/\eta) \geq \eta^*_{R,\mathcal{P}}\mathbf{C}(\mathcal{P}/\eta^*_{R,\mathcal{P}})$. Since $\eta^*_{R,\mathcal{P}}\mathbf{C}(\mathcal{P}/\eta^*_{R,\mathcal{P}}) = R$, we have $\eta\mathbf{C}(P/\eta) \geq R$. Since $\mathbf{C}(\mathcal{P}_1)$ is strictly increasing for $\mathcal{P}_1 \leq \mathcal{P}^*$, this shows that $\mathcal{P}_1 \leq \mathcal{P}/\eta$ for $\mathcal{P}_1 \leq \mathcal{P}^*$.

Finally, consider $\eta \notin \mathcal{I}_{R,\mathcal{P}}$. If $\eta < R/\mathbf{C}^*$, then no solution exists for $R = \eta\mathbf{C}(P_1)$. If $\frac{R}{\mathbf{C}^*} \leq \eta < \eta^*_{R,\mathcal{P}}$, then the strict monotonicity of $\eta\mathbf{C}(\mathcal{P}/\eta)$ in this range shows that $\eta\mathbf{C}(\mathcal{P}/\eta) < \eta^*_{R,\mathcal{P}}\mathbf{C}(\mathcal{P}/\eta^*_{R,\mathcal{P}}) = R$. For $\mathcal{P}_1$ satisfying $R = \eta\mathbf{C}(\mathcal{P}_1)$, then, $\eta\mathbf{C}(\mathcal{P}/\eta) < \eta\mathbf{C}(\mathcal{P}_1)$ and the strict monotonicity of $\mathbf{C}(\cdot)$ shows that $\mathcal{P}/\eta < \mathcal{P}_1$, so the condition on $\mathcal{P}_2$ must be violated. Similarly if $\eta \frac{R}{\mathbf{C}(0)}$ then $\forall\mathcal{P}_1 \geq 0$, $\mathbf{C}(\mathcal{P}_1) > R$, which will violate the rate condition. **QED**

**Proof of Lemma 5 (Concavity of $E(R, \mathcal{P}, \eta)$):**

For any given DMC, let $\Omega$ be the set of triples $(R, \mathcal{P}, \eta)$ for which $0 < R < \mathbf{C}(\mathcal{P})$ and $\eta \in \mathcal{I}_{R,\mathcal{P}}$.

$$\Omega = \{(R, \mathcal{P}, \eta) : \mathcal{P} \geq 0, 0 < R < \mathbf{C}(\mathcal{P}), \eta \in \mathcal{I}_{R,\mathcal{P}}\}$$

First we show that $\Omega$ is a convex set, and then we show that $E(R, \mathcal{P}, \eta)$ is a concave function over the domain $\Omega$.

Assume that $(R', \mathcal{P}', \eta')$ and $(R'', \mathcal{P}'', \eta'')$ are arbitrary points of $\Omega$. We show that $\Omega$ is a convex set by showing that for any $\alpha \in (0, 1)$, the point $(R_\alpha, \mathcal{P}_\alpha, \eta_\alpha)$ given by

$$\mathcal{P}_\alpha = \alpha\mathcal{P}' + (1-\alpha)\mathcal{P}'' \qquad R_\alpha = \alpha R' + (1-\alpha)R'' \qquad \eta_\alpha = \alpha\eta' + (1-\alpha)\eta'' \qquad (45)$$

is also in the set $\Omega$.

$R_\alpha$ is clearly positive, and using the concavity of $\mathbf{C}(\cdot)$, we get

$$R_\alpha < \alpha\mathbf{C}(\mathcal{P}') + (1-\alpha)\mathbf{C}(\mathcal{P}'') \leq \mathbf{C}(\alpha\mathcal{P}' + (1-\alpha)\mathcal{P}'') = \mathbf{C}(\mathcal{P}_\alpha)$$

We must also show that $\eta_\alpha \in \mathcal{I}_{R_\alpha, \mathcal{P}_\alpha}$. It suffices to show that $\eta_\alpha \geq \eta^*_{R_\alpha, \mathcal{P}_\alpha}$, $\eta_\alpha < 1$ and $\eta_\alpha \leq R_\alpha/C(0)$. The latter two conditions are obvious, so we must show only that $\eta_\alpha \geq \eta^*_{R_\alpha, \mathcal{P}_\alpha}$. As shown in the proof of Lemma 4, the condition $\eta \geq \eta^*_{R,\mathcal{P}}$ is equivalent to $R \leq \eta\mathbf{C}(\mathcal{P}/\eta)$.

$$\begin{aligned}R' \leq \eta'\mathbf{C}\left(\frac{\mathcal{P}'}{\eta'}\right) \\ R'' \leq \eta''\mathbf{C}\left(\frac{\mathcal{P}''}{\eta''}\right)\end{aligned} \Rightarrow R_\alpha \leq \alpha\eta'\mathbf{C}\left(\frac{\mathcal{P}'}{\eta'}\right) + (1-\alpha)\mathbf{C}\left(\frac{\mathcal{P}''}{\eta''}\right)$$

---

[23] Note that for $\eta = R/\mathcal{P}^*$, the equation $R = \eta\mathbf{C}(\mathcal{P}_1)$ is also satisfied for all $\mathcal{P}_1 > \mathcal{P}^*$. These values of $\mathcal{P}_1$ are omitted from the optimization since they can be reduced to $\mathcal{P}^*$ thus allowing more energy for phase 2.

Thus,
$$\begin{aligned}
R_\alpha &\leq \eta_\alpha \left( \frac{\alpha \eta'}{\eta_\alpha} \mathbf{C}\left(\frac{\mathcal{P}'}{\eta'}\right) + \frac{(1-\alpha)\eta''}{\eta_\alpha} \mathbf{C}\left(\frac{\mathcal{P}''}{\eta''}\right) \right) \\
&\leq \eta_\alpha \mathbf{C}\left( \frac{\alpha \eta'}{\eta_\alpha} \frac{\mathcal{P}'}{\eta'} + \frac{(1-\alpha)\eta''}{\eta_\alpha} \frac{\mathcal{P}''}{\eta''} \right) \\
&= \eta_\alpha \mathbf{C}\left( \frac{\alpha \mathcal{P}' + (1-\alpha)\mathcal{P}''}{\eta_\alpha} \right) \\
&= \eta_\alpha \mathbf{C}\left( \frac{\mathcal{P}_\alpha}{\eta_\alpha} \right).
\end{aligned}$$

Consequently $\Omega$ is a convex region. We next show that $E(R, \mathcal{P}, \eta)$ is concave over $\Omega$. That is, given points $(R', \mathcal{P}', \eta'), (R'', \mathcal{P}'', \eta'')$ and $(R_\alpha, \mathcal{P}_\alpha, \eta_\alpha)$ in $\Omega$, we will show that $L_\alpha \leq E(R_\alpha, \mathcal{P}_\alpha, \eta_\alpha)$ where $L_\alpha = \alpha E(R', \mathcal{P}', \eta') + (1-\alpha) E(R'', \mathcal{P}'', \eta'')$.

$$\begin{aligned}
L_\alpha &= \alpha(1-\eta')\mathbf{D}(\mathcal{P}'_2) + (1-\alpha)(1-\eta'')\mathbf{D}(\mathcal{P}''_2) \\
&\leq (1-\eta_\alpha)\mathbf{D}\left( \frac{\alpha(1-\eta')\mathcal{P}'_2}{1-\eta_\alpha} + \frac{(1-\alpha)(1-\eta'')\mathcal{P}''_2}{1-\eta_\alpha} \right) \\
&= (1-\eta_\alpha)\mathbf{D}\left( \frac{\alpha\left(\mathcal{P}' - \mathbf{C}^{-1}\left(\frac{R'}{\eta'}\right)\right)}{1-\eta_\alpha} + \frac{(1-\alpha)\left(\mathcal{P}'' - \mathbf{C}^{-1}\left(\frac{R''}{\eta''}\right)\right)}{1-\eta_\alpha} \right) \\
&= \eta_\alpha \mathbf{D}\left( \frac{\mathcal{P}_\alpha - \left[\alpha\eta'\mathbf{C}^{-1}\left(\frac{R'}{\eta'}\right) + (1-\alpha)\eta''\mathbf{C}^{-1}\left(\frac{R''}{\eta''}\right)\right]}{1-\eta_\alpha} \right) \\
&\leq \eta_\alpha \mathbf{D}\left( \frac{\mathcal{P}_\alpha - \eta_\alpha \mathbf{C}^{-1}\left(\frac{\alpha R' + (1-\alpha)R''}{\eta_\alpha}\right)}{1-\eta_\alpha} \right) = E(R_\alpha, \mathcal{P}_\alpha, \eta_\alpha).
\end{aligned}$$

The first inequality above used the concavity of $\mathbf{D}(\cdot)$ combined with $1-\eta_\alpha = \alpha(1-\eta')+(1-\alpha)(1-\eta'')$. The second inequality used the convexity of $\mathbf{C}^{-1}(\cdot)$ along with the fact that $\mathbf{D}(\cdot)$ is non-decreasing.
**QED**

**Proof of Lemma 6:**

We will first prove that $\mathbf{E}[|\mathbf{V}_n|] < \infty$. Recall

$$\mathbf{V}_n = \mathcal{H}_{\mathcal{F}_n} + n\mathbf{C}(\mathcal{P}) + \gamma_\mathbf{C}^\mathcal{P} \left( \mathbf{E}\left[\mathcal{S}_n | \mathcal{F}_n\right] - \mathcal{P}n \right)$$

Since $\gamma_\mathbf{C}^\mathcal{P} \geq 0$, $\mathbf{C}(\mathcal{P}) \geq 0$ and $\mathbf{E}[\mathcal{S}_n | \mathcal{F}_n] \geq 0$,

$$|\mathbf{V}_n| \leq |\mathcal{H}_{\mathcal{F}_n}| + n\mathbf{C}(\mathcal{P}) + \gamma_\mathbf{C}^\mathcal{P} \left( \mathbf{E}\left[\mathcal{S}_n | \mathcal{F}_n\right] + \mathcal{P}n \right)$$

Note that $|\mathcal{H}_{\mathcal{F}_n}| \leq \ln M$.

$$\mathbf{E}[|\mathbf{V}_n|] \leq \ln M + n\mathbf{C}(\mathcal{P}) + \gamma_\mathbf{C}^\mathcal{P} \left( \mathbf{E}\left[\mathcal{S}_n\right] + \mathcal{P}n \right)$$

In addition for any finite energy code[24] $\mathbf{E}[\mathcal{S}_n] < \infty$. Consequently $\mathbf{E}[|\mathbf{V}_n|] < \infty$

---

[24]The convention for extending the encoding algorithms beyond decoding time is assigning all of the codewords to the same zero cost symbol.

We next prove that $\mathbf{E}\left[\mathbf{V}_{n+1} | \mathcal{F}_n\right] \geq \mathbf{V}_n$.

$$\mathbf{E}\left[\mathbf{V}_n | \mathcal{F}_n\right] = \mathbf{E}\left[\mathcal{H}_{\mathcal{F}_n} + n\mathbf{C}(\mathcal{P}) + \gamma_{\mathbf{C}}^{\mathcal{P}}(\mathcal{S}_n - n\mathcal{P}) | \mathcal{F}_n\right]$$

$$\mathbf{E}\left[\mathbf{V}_{n+1} | \mathcal{F}_n\right] = \mathbf{E}\left[\mathcal{H}_{\mathcal{F}_{n+1}} + (n+1)\mathbf{C}(\mathcal{P}) + \gamma_{\mathbf{C}}^{\mathcal{P}}(\mathcal{S}_n + \rho_{X_{n+1}} - (n+1)\mathcal{P}) \Big| \mathcal{F}_n\right]$$

$$\mathbf{E}\left[\mathbf{V}_{n+1} | \mathcal{F}_n\right] = \mathbf{V}_n + \mathbf{C}(\mathcal{P}) - \gamma_{\mathbf{C}}^{\mathcal{P}}\mathcal{P} - I(\theta; Y_{n+1} | \mathcal{F}_n = \mathfrak{f}_n) + \mathbf{E}\left[\rho_{X_{n+1}} \Big| \mathcal{F}_n\right]$$

Because of the Markov relations, $\theta \leftrightarrow X_{n+1} \leftrightarrow Y_{n+1}$ which holds for all $\mathfrak{f}_n$ combined with the data processing inequality, we have

$$\mathbf{E}\left[\mathbf{V}_{n+1} | \mathcal{F}_n\right] \geq \mathbf{V}_n + \mathbf{C}(\mathcal{P}) - \gamma_{\mathbf{C}}^{\mathcal{P}}\mathcal{P} - I(X_{n+1}; Y_{n+1} | \mathcal{F}_n = \mathfrak{f}_n) + \gamma_{\mathbf{C}}^{\mathcal{P}}\mathbf{E}\left[\rho_{X_{n+1}} \Big| \mathcal{F}_n\right]$$

Note that

$$\mathbf{C}(\mathcal{P}) - \gamma_{\mathbf{C}}^{\mathcal{P}}\mathcal{P} = \max_{\varphi}\left(\Im(\varphi) - \gamma_{\mathbf{C}}^{\mathcal{P}}\sum_k \varphi(k)\rho_k\right),$$

where $\Im(\varphi)$ is the mutual information corresponding to the input distribution $\varphi$. Thus

$$\mathbf{E}\left[\mathbf{V}_{n+1} | \mathcal{F}_n\right] \geq \mathbf{V}_n$$

and the stochastic sequence $\{\mathbf{V}_n, n\}$ is a submartingale.  **QED**

**Proof of Lemma 7:**

We use the following shorthand for a given $\mathfrak{f}_n$:

$$p(i) = p_i(\mathfrak{f}_n) \qquad\qquad p(i|j) = \mathbf{P}\left[\theta = i | Y_{n+1} = j, \mathcal{F}_n = \mathfrak{f}_n\right]$$
$$\varphi(k|i) = \mathbf{P}\left[X_{n+1} = k | \mathcal{F}_n = \mathfrak{f}_n, \theta = i\right] \qquad\qquad \varphi(k) = \mathbf{P}\left[X_{n+1} = k | \mathcal{F}_n = \mathfrak{f}_n\right]$$
$$\psi(j|i) = \mathbf{P}\left[Y_{n+1} = j | \mathcal{F}_n = \mathfrak{f}_n, \theta = i\right] \qquad\qquad \psi(j) = \mathbf{P}\left[Y_{n+1} = j | \mathcal{F}_n = \mathfrak{f}_n\right]$$

We proceed to upper bound $\mathbf{E}\left[\ln \mathcal{H}_{\mathcal{F}_n} - \ln \mathcal{H}_{\mathcal{F}_{n+1}} \Big| \mathcal{F}_n = \mathfrak{f}_n\right]$.

$$\mathbf{E}\left[\ln \mathcal{H}_{\mathcal{F}_n} - \ln \mathcal{H}_{\mathcal{F}_{n+1}} \Big| \mathcal{F}_n = \mathfrak{f}_n\right] = \sum_{j=1}^{|\mathcal{Y}|} \psi(j) \ln \frac{\sum_{i=1}^M p(i) \ln \frac{1}{p(i)}}{\sum_{i=1}^M p(i|j) \ln \frac{1}{p(i|j)}}$$

$$\leq \sum_i \frac{-p(i) \ln p(i)}{\sum_m -p(m) \ln p(m)} \sum_j \psi(j) \ln \frac{p(i) \ln \frac{1}{p(i)}}{p(i|j) \ln \frac{1}{p(i|j)}}$$

where we have used the log-sum inequality for each $j$ above.

Using $\frac{p(i)}{p(i|j)} = \frac{\psi(j)}{\psi(j|i)}$ and defining $\psi(j|\bar{i}) \triangleq \mathbf{P}\left[Y_{n+1}{=}j | \mathcal{F}_n{=}\mathfrak{f}_n, \theta{\neq}i\right]$,

$$\mathbf{E}\left[\ln \mathcal{H}_{\mathcal{F}_n} - \ln \mathcal{H}_{\mathcal{F}_{n+1}} \Big| \mathcal{F}_n = \mathfrak{f}_n\right] = \sum_{j=1}^{|\mathcal{Y}|} \psi(j) \ln \frac{\sum_{i=1}^M p(i) \ln \frac{1}{p(i)}}{\sum_{i=1}^M p(i|j) \ln \frac{1}{p(i|j)}}$$

$$\leq \sum_i \frac{p(i) \ln \frac{1}{p(i)}}{\sum_m p(m) \ln \frac{1}{p(m)}} \sum_j \psi(j) \ln \frac{p(i) \ln \frac{1}{p(i)}}{p(i|j) \ln \frac{1}{p(i|j)}}$$

where we have used the log-sum inequality for each $j$ above.

Using $\frac{p(i)}{p(i|j)} = \frac{\psi(j)}{\psi(j|i)}$ and defining $\psi(j|\bar{i}) = \mathbf{P}\left[Y_{n+1}=j \mid \mathcal{F}_n=\mathfrak{f}_n, \theta \neq i\right]$,

$$\sum_j \psi(j) \ln \frac{p(i) \ln p(i)}{p(i|j) \ln p(i|j)} = \sum_j \psi(j) \ln \left( \frac{\psi(j)}{\psi(j|i)} \frac{\ln p(i)}{\ln p(i|j)} \right)$$

$$= \sum_j [p(i)\psi(j|i) + (1-p(i))\psi(j|\bar{i})] \ln \left( \frac{\psi(j)}{\psi(j|i)} \frac{\ln p(i)}{\ln p(i|j)} \right)$$

$$\leq p(i) \sum_j \psi(j|i) \ln \frac{\psi(j|i)}{\psi(j|\bar{i})} + (1-p(i)) \sum_j \psi(j|\bar{i}) \ln \frac{\psi(j|\bar{i})}{\psi(j|i)} \quad (46)$$

In order to verify the inequality in (46), denote the right side minus the left side as $A$, and note that by substitution,

$$A = p(i) \sum_j \psi(j|i) \ln \frac{\psi(j|i)}{\psi(j)} + (1-p(i)) \sum_j \psi(j|\bar{i}) \ln \frac{\psi(j|\bar{i})}{\psi(j)}$$

$$+ p(i) \sum_j \psi(j|i) \ln \left( \frac{\psi(j|i)}{\psi(j|\bar{i})} \frac{\ln p(i|j)}{\ln p(i)} \right) + (1-p(i)) \sum_j \psi(j|\bar{i}) \ln \left( \frac{\ln p(i|j)}{\ln p(i)} \right)$$

The first two terms above are divergences, and thus non-negative. The third term can be rewritten as below and is shown to be nonnegative by applying Jensen's inequality to the function $\ln\left(\frac{1}{x}\ln(1+\alpha x)\right)$ which is convex for any $\alpha > 0$.

$$\sum_j \psi(j|i) \ln \left( \frac{\psi(j|i)}{\psi(j|\bar{i})} \frac{\ln p(i|j)}{\ln p(i)} \right) = \sum_j \psi(j|i) \ln \frac{\psi(j|i)}{\psi(j|\bar{i})} \frac{\ln\left(1 + \frac{1-p(i)}{p(i)} \frac{\psi(j|\bar{i})}{\psi(j|i)}\right)}{\ln\left(1 + \frac{1-p(i)}{p(i)}\right)} \geq 0$$

Similarly, the fourth term can be rewritten as below and is shown to be nonnegative by applying Jensen's inequality to the convex function $\ln\left(\ln(1+\frac{1}{\alpha x})\right)$ for $\alpha > 0$.

$$\sum_j \psi(j|\bar{i}) \ln \left( \frac{\ln p(i|j)}{\ln p(i)} \right) = \sum_j \psi(j|\bar{i}) \ln \frac{\ln\left(1 + \frac{1-p(i)}{p(i)} \frac{\psi(j|\bar{i})}{\psi(j|i)}\right)}{\ln\left(1 + \frac{1-p(i)}{p(i)}\right)} \geq 0$$

This verifies (46). The final term in (46) can be upper bounded by

$$\sum_j \psi(j|\bar{i}) \ln \frac{\psi(j|\bar{i})}{\psi(j|i)} = \sum_j \left( \sum_k \varphi(k|\bar{i}) P_{kj} \right) \ln \frac{\sum_k \varphi(k|\bar{i}) P_{kj}}{\psi(j|i) \sum_k \varphi(k|\bar{i})}$$

$$\stackrel{(a)}{\leq} \sum_j \sum_k \varphi(k|\bar{i}) P_{kj} \ln \frac{P_{kj}}{\psi(j|i)}$$

$$\stackrel{(b)}{\leq} \sum_k \varphi(k|\bar{i}) \mathbf{D}_k.$$

where $(a)$ uses the log sum inequality over $k$ for each $j$ and $(b)$ follows from the definition of $\mathbf{D}_k$.

By a similar argument on the first term in (46),

$$\sum_j \psi(j|i) \ln \frac{\psi(j|i)}{\psi(j|\hat{i})} \leq \sum_k \varphi(k|i) \mathbf{D}_k.$$

Substituting the above two inequalities into (46),

$$\sum_j \psi(j) \ln \frac{p(i) \ln p(i)}{p(i|j) \ln p(i|j)} \leq \sum_k \varphi(k) \mathbf{D}_k.$$

Thus

$$\mathbf{E}\left[\ln \mathcal{H}_{\mathcal{F}_n} - \ln \mathcal{H}_{\mathcal{F}_{n+1}} \middle| \mathcal{F}_n = \mathfrak{f}_n\right] \leq \sum_k \varphi(k) \mathbf{D}_k = \mathbf{E}\left[\mathbf{D}_{X_{n+1}} \middle| \mathcal{F}_n = \mathfrak{f}_n\right], \qquad (47)$$

which is equivalent to (23). **QED**

**Proof of Lemma 8:**

We will first prove that $\mathbf{E}[\mathbf{W}_{n+1}|\mathcal{F}_n] \geq \mathbf{W}_n$. Recalling that $\mathbf{W}_n = \ln \mathcal{H}_{\mathcal{F}_n} + n\mathbf{D}(\mathcal{P}) + \gamma_{\mathbf{D}}^{\mathcal{P}}(\mathbf{E}[\mathcal{S}_n|\mathcal{F}_n] - n\mathcal{P})$ and using Lemma 7,

$$\mathbf{E}[\mathbf{W}_{n+1}|\mathcal{F}_n] \geq \mathbf{W}_n + \mathbf{D}(\mathcal{P}) - \mathbf{E}[\mathbf{D}_{X_{n+1}}|\mathcal{F}_n] - \gamma_{\mathbf{D}}^{\mathcal{P}}\left(\mathbf{E}[\rho_{X_{n+1}}|\mathcal{F}_n] - \mathcal{P}\right)$$
$$\geq \mathbf{W}_n$$

We next prove that $\mathbf{E}[|\mathbf{W}_n|] < \infty$. Using the definition of $\mathbf{W}_n$ and the fact that $\gamma_{\mathbf{D}}^{\mathcal{P}} \geq 0$, $\mathbf{D}(\mathcal{P}) \geq 0$ and $\mathbf{E}[\mathcal{S}_n|\mathcal{F}_n] \geq 0$,

$$|\mathbf{W}_n| \leq |\ln \mathcal{H}_{\mathcal{F}_n}| + n\mathbf{D}(\mathcal{P}) + \gamma_{\mathbf{D}}^{\mathcal{P}}\left(\mathbf{E}[\mathcal{S}_n|\mathcal{F}_n] + \mathcal{P}n\right)$$

Note that since $\mathcal{H}_{\mathcal{F}_n} \leq \mathcal{H}_{\mathcal{F}_0} = \ln M$.

$$\mathbf{E}[|\mathbf{W}_n|] \leq \ln M + \mathbf{E}\left[\ln \frac{\mathcal{H}_{\mathcal{F}_0}}{\mathcal{H}_{\mathcal{F}_n}}\right] + n\mathbf{D}(\mathcal{P}) + \gamma_{\mathbf{D}}^{\mathcal{P}}\left(\mathbf{E}[\mathcal{S}_n] + \mathcal{P}n\right)$$

Since for any finite energy code[25] $\mathbf{E}[\mathcal{S}_n] < \infty$, proving that $\mathbf{E}\left[\ln \frac{\mathcal{H}_{\mathcal{F}_0}}{\mathcal{H}_{\mathcal{F}_n}}\right] < \infty$, will prove $\mathbf{E}[|\mathbf{W}_n|] < \infty$.

Note that

$$\mathbf{E}\left[\ln \frac{\mathcal{H}_{\mathcal{F}_0}}{\mathcal{H}_{\mathcal{F}_n}}\right] = \mathbf{E}\left[\sum_{k=1}^n \ln \frac{\mathcal{H}_{\mathcal{F}_{k-1}}}{\mathcal{H}_{\mathcal{F}_k}}\right]$$

Using the convexity of $\mathbf{D}(\mathcal{P})$ function together with equation (47) we get

$$\mathbf{E}\left[\sum_{k=1}^n \ln \frac{\mathcal{H}_{\mathcal{F}_{k-1}}}{\mathcal{H}_{\mathcal{F}_k}}\right] \leq n\mathbf{E}\left[\mathbf{D}\left(\frac{\mathbf{E}[\mathcal{S}_n]}{n}\right)\right]$$

Recalling that $\mathbf{E}[\mathcal{S}_n] < \infty$, will prove $\mathbf{E}\left[\ln \frac{\mathcal{H}_{\mathcal{F}_0}}{\mathcal{H}_{\mathcal{F}_n}}\right] < \infty$ and thus $\mathbf{E}[|W_n^{\mathcal{P}}|] < \infty$. **QED**

**Proof of Lemma 9:**

---

[25] Recall the convention of extending the encoding algorithm beyond decoding time by assigning all of the codewords to the same zero cost symbol.

By the definition of $\mathbf{V}_n$,

$$\mathbf{V}_{\tau_i} = \mathcal{H}_{\mathcal{F}_{\tau_i}} + \tau_i \mathbf{C}(\mathcal{P}) + \gamma_{\mathbf{C}}^{\mathcal{P}} \left( \mathbf{E}\left[ \mathcal{S}_{\tau_i} | \mathcal{F}_{\tau_i} \right] - \mathcal{P}\tau_i \right) \tag{48}$$

Since the expected value of each term on the right side exists,

$$\begin{aligned}
\mathbf{E}\left[\mathbf{V}_{\tau_i}^{\mathcal{P}}\right] &= \mathbf{E}\left[\mathcal{H}_{\mathcal{F}_{\tau_i}}\right] + \mathbf{E}\left[\tau_i\right] \mathbf{C}(\mathcal{P}) + \gamma_{\mathbf{C}}^{\mathcal{P}} \mathbf{E}\left[\mathcal{S}_{\tau_i} - \mathcal{P}\tau_i\right] \\
&\leq \mathbf{E}\left[\mathcal{H}_{\mathcal{F}_{\tau_i}}\right] + \mathbf{E}\left[\tau_i\right] \mathbf{C}(\mathcal{P}),
\end{aligned} \tag{49}$$

where we have used $\gamma_{\mathbf{C}}^{\mathcal{P}} \geq 0$ along with the hypothesis of the lemma that $\mathbf{E}\left[\mathcal{S}_{\tau_i}\right] \leq \mathcal{P} \mathbf{E}\left[\tau_i\right]$. Since $\mathbf{E}\left[V_0^{\mathcal{P}}\right] = \mathbf{E}\left[\mathcal{H}_{\mathcal{F}_0}\right] = \ln M$, the result of the lemma, i.e., $\mathbf{C}(\mathcal{P})\mathbf{E}\left[\tau_i\right] \geq \mathbf{E}\left[\mathcal{H}_{\mathcal{F}_0}\right] - \mathbf{E}\left[\mathcal{H}_{\mathcal{F}_{\tau_i}}\right]$ will then hold if

$$\mathbf{E}\left[\mathbf{V}_{\tau_i}\right] \geq \mathbf{E}\left[\mathbf{V}_0\right] \tag{50}$$

holds. Doob's theorem[26] states that a submartingale $\mathbf{V}_n$ satisfies (50) if it satisfies the following two conditions:

$$\mathbf{E}\left[|\mathbf{V}_{\tau_i}|\right] < \infty \text{ and } \lim_{n \to \infty} \mathbf{E}\left[|\mathbf{V}_n| \mathbb{I}_{\{\tau_i \geq n\}}\right] = 0 \tag{51}$$

The first condition follows from modifying (48) to bound $|\mathbf{V}_{\tau_i}^{\mathcal{P}}|$.

$$|\mathbf{V}_{\tau_i}^{\mathcal{P}}| \leq \mathcal{H}_{\mathcal{F}_{\tau_i}} + \tau_i \mathbf{C}(\mathcal{P}) + \gamma_{\mathbf{C}}^{\mathcal{P}} \left( \mathbf{E}\left[\mathcal{S}_{\tau_i} | \mathcal{F}_{\tau_i}\right] + \mathcal{P}\tau_i \right)$$

To establish the second condition, let

$$\begin{aligned}
\xi_n &= |\mathbf{V}_n| \mathbb{I}_{\{\tau_i \geq n\}} \tag{52} \\
&= |\mathcal{H}_{\mathcal{F}_n} + n\mathbf{C}(\mathcal{P}) + \gamma_{\mathbf{C}}^{\mathcal{P}} \mathbf{E}\left[\mathcal{S}_n | \mathcal{F}_n\right] - \gamma_{\mathbf{C}}^{\mathcal{P}} \mathcal{P} n| \mathbb{I}_{\{\tau_i \geq n\}} \\
&\leq \left[\mathcal{H}_{\mathcal{F}_n} + n\mathbf{C}(\mathcal{P}) + \gamma_{\mathbf{C}}^{\mathcal{P}} \mathbf{E}\left[\mathcal{S}_n | \mathcal{F}_n\right] + \gamma_{\mathbf{C}}^{\mathcal{P}} \mathcal{P} n\right] \mathbb{I}_{\{\tau_i \geq n\}} \tag{53}
\end{aligned}$$

We want to find a random variable $\zeta$ of finite expectation that upper bounds $\xi_n$ for each $n$; the troublesome term here is $\mathbf{E}\left[\mathcal{S}_n | \mathcal{F}_n\right]$. Let $\mathcal{S}_n(m)$ (which is measurable in $\mathcal{F}_n$) be the cost of the codeword corresponding to the message $m$ at time $n$. The following very weak bound is sufficient for our purposes.

$$\mathbf{E}\left[\mathcal{S}_n | \mathcal{F}_n\right] = \sum_{m=1}^{M} \mathbf{P}\left[\theta{=}m | \mathcal{F}_n\right] \mathcal{S}_n(m) \leq \sum_{m=1}^{M} \mathcal{S}_n(m) \tag{54}$$

Substituting (54) into (53),

$$\begin{aligned}
\xi_n &\leq \left[\ln M + n\mathbf{C}(\mathcal{P}) + \gamma_{\mathbf{C}}^{\mathcal{P}} \sum_{m=1}^{M} \mathcal{S}_n(m) + \gamma_{\mathbf{C}}^{\mathcal{P}} \mathcal{P} n\right] \mathbb{I}_{\{\tau_i \geq n\}} \\
&\stackrel{(a)}{\leq} \left[\ln M + \tau_i \mathbf{C}(\mathcal{P}) + \gamma_{\mathbf{C}}^{\mathcal{P}} \sum_{m=1}^{M} \mathcal{S}_{\tau_i}(m) + \gamma_{\mathbf{C}}^{\mathcal{P}} \mathcal{P}\tau_i\right] \mathbb{I}_{\{\tau_i \geq n\}} \\
&\leq \left[\ln M + \tau_i \mathbf{C}(\mathcal{P}) + \gamma_{\mathbf{C}}^{\mathcal{P}} \sum_{m=1}^{M} \mathcal{S}_{\tau_i}(m) + \gamma_{\mathbf{C}}^{\mathcal{P}} \mathcal{P}\tau_i\right] \\
&\leq \left[\ln M + \tau_i \mathbf{C}(\mathcal{P}) + \gamma_{\mathbf{C}}^{\mathcal{P}} M \mathbf{E}\left[\mathcal{S}_{\tau_i}\right] + \gamma_{\mathbf{C}}^{\mathcal{P}} \mathcal{P}\tau_i\right] \triangleq \zeta \tag{55}
\end{aligned}$$

---

[26]See, for example, Shiryaev, [17], page 457.

In $(a)$ we have used the fact that indicator function is zero if $\tau_i < n$; and $\mathcal{S}_{\tau_i}(m) \geq \mathcal{S}_n(m)$ if $\tau_i > n$. Note that $0 \leq \xi_n \leq \zeta$ for all $n$. Since $\mathbf{E}[\tau_i] < \infty$ and $\mathbf{E}[\mathcal{S}_{\tau_i}] \leq \mathcal{P}\mathbf{E}[\tau_i] < \infty$, $\mathbf{E}[\zeta] < \infty$. Since $\lim_{n \to \infty} \mathbf{P}[\xi_n = 0] = 1$, Lebesgue's dominated convergence theorem shows that $\lim_{n \to \infty} \mathbf{E}[\xi_n] = 0$.

**QED**

**Proof of Lemma 10:**

Lemma 8 showed that the sequence

$$\mathbf{W}_n = \ln \mathcal{H}_{\mathcal{F}_n} + n\mathbf{D}(\mathcal{P}) + \gamma_{\mathbf{D}}^{\mathcal{P}}(\mathbf{E}[\mathcal{S}_n | \mathcal{F}_n] - n\mathcal{P}) \tag{56}$$

for $n \geq 0$ is a submartingale, and we will use Doob's theorem to prove the lemma. In particular, for two stopping times, $\tau_i \leq \tau_f$, Doob's theorem says that if, for both $s = i$ and $s = f$,

$$\mathbf{E}[|\mathbf{W}_{\tau_s}|] < \infty \quad \text{and} \quad \lim_{n \to \infty} \mathbf{E}[|\mathbf{W}_n| \mathbb{I}_{\{\tau_s \geq n\}}] = 0 \tag{57}$$

then $\mathbf{E}[\mathbf{W}_{\tau_i}]$ and $\mathbf{E}[\mathbf{W}_{\tau_f}]$ exist and satisfy

$$\mathbf{E}[\mathbf{W}_{\tau_f}] \geq \mathbf{E}[\mathbf{W}_{\tau_i}] \tag{58}$$

For the moment, assume that the condition of (57) is satisfied. Then substituting the definition of $\mathbf{W}_n$ for $n = \tau_i$ and $n = \tau_f$ into (58),

$$\mathbf{E}\left[\ln \frac{\mathcal{H}_{\mathcal{F}_{\tau_f}}}{\mathcal{H}_{\mathcal{F}_{\tau_i}}}\right] + \mathbf{D}(\mathcal{P})\mathbf{E}[\tau_f - \tau_i] + \gamma_{\mathbf{D}}^{\mathcal{P}}\left(\mathbf{E}[\mathcal{S}_{\tau_f} - \mathcal{S}_{\tau_i}] - \mathcal{P}\mathbf{E}[\tau_f - \tau_i]\right) \geq 0$$

Inserting the assumption $\mathbf{E}[\mathcal{S}_{\tau_f} - \mathcal{S}_{\tau_i}] \leq \mathcal{P}\mathbf{E}[\tau_f - \tau_i]$,

$$\mathbf{E}\left[\ln \frac{\mathcal{H}_{\mathcal{F}_{\tau_f}}}{\mathcal{H}_{\mathcal{F}_{\tau_i}}}\right] + \mathbf{D}(\mathcal{P})\mathbf{E}[\tau_f - \tau_i] \geq 0$$

This is equivalent to the result of the lemma, so we need only establish the condition in (57) to complete the proof. For the first part, we can modify (56) to bound $|\mathbf{W}_{\tau_s}|$ as

$$|\mathbf{W}_{\tau_s}| \leq |\ln \mathcal{H}_{\mathcal{F}_{\tau_s}}| + \tau_s \mathbf{D}(\mathcal{P}) + \gamma_{\mathbf{D}}^{\mathcal{P}}(\mathbf{E}[\mathcal{S}_{\tau_s} | \mathcal{F}_{\tau_s}] + \tau_s \mathcal{P})$$

All but the first of these terms clearly have finite expectations, so the first part of (57) reduces to proving that $\mathbf{E}\left[|\ln \mathcal{H}_{\mathcal{F}_{\tau_s}}|\right] < \infty$. Since $\mathcal{H}_{\mathcal{F}_0} = \ln M$,

$$|\ln \mathcal{H}_{\mathcal{F}_{\tau_s}}| = \left|\sum_{n=0}^{\tau_s - 1} \ln \frac{\mathcal{H}_{\mathcal{F}_{n+1}}}{\mathcal{H}_{\mathcal{F}_n}} + \ln \ln M\right|$$

$$\leq \sum_{n=1}^{\tau_s - 1} \left|\ln \frac{\mathcal{H}_{\mathcal{F}_n}}{\mathcal{H}_{\mathcal{F}_{n+1}}}\right| + |\ln \ln M| \tag{59}$$

Now using Lemma 7, we have

$$\mathbf{E}\left[\ln \frac{\mathcal{H}_{\mathcal{F}_n}}{\mathcal{H}_{\mathcal{F}_{n+1}}} \bigg| \mathcal{F}_n\right] \leq \mathbf{E}\left[\mathbf{D}_{X_{n+1}} \big| \mathcal{F}_n\right] \tag{60}$$

$$\leq \mathbf{D}\left(\mathbf{E}[\mathcal{S}_{n+1} - \mathcal{S}_n | \mathcal{F}_n]\right) \tag{61}$$

where the second inequality follows from the concavity of the function $\mathbf{D}(\cdot)$.

For any random variable $v$, we have $v = v\mathbb{I}_{\{v \geq 0\}} + v\mathbb{I}_{\{v < 0\}}$ and $|v| = v\mathbb{I}_{\{v \geq 0\}} - v\mathbb{I}_{\{v < 0\}}$. Combining these, $|v| = v - 2v\mathbb{I}_{\{v < 0\}}$. Applying this to the random variable $\ln \mathcal{H}_{\mathcal{F}_n} - \ln \mathcal{H}_{\mathcal{F}_{n+1}}$,

$$\mathbf{E}\left[\left|\ln \frac{\mathcal{H}_{\mathcal{F}_n}}{\mathcal{H}_{\mathcal{F}_{n+1}}}\right| \,\bigg|\, \mathcal{F}_n\right] \leq \mathbf{D}\left(\mathbf{E}\left[\mathcal{S}_{n+1} - \mathcal{S}_n|\mathcal{F}_n\right]\right) - 2\mathbf{E}\left[\ln \frac{\mathcal{H}_{\mathcal{F}_n}}{\mathcal{H}_{\mathcal{F}_{n+1}}} \mathbb{I}_{\{\mathcal{H}_{\mathcal{F}_n} < \mathcal{H}_{\mathcal{F}_{n+1}}\}} \,\bigg|\, \mathcal{F}_n\right]$$

The last term above can be bounded as

$$\begin{aligned}
\mathbf{E}\left[\ln \frac{\mathcal{H}_{\mathcal{F}_{n+1}}}{\mathcal{H}_{\mathcal{F}_n}} \mathbb{I}_{\{\mathcal{H}_{\mathcal{F}_n} \leq \mathcal{H}_{\mathcal{F}_{n+1}}\}} \,\bigg|\, \mathcal{F}_n\right] &= \mathbf{E}\left[\frac{\mathcal{H}_{\mathcal{F}_{n+1}}}{\mathcal{H}_{\mathcal{F}_n}} \left(-\frac{\mathcal{H}_{\mathcal{F}_n}}{\mathcal{H}_{\mathcal{F}_{n+1}}} \ln \frac{\mathcal{H}_{\mathcal{F}_n}}{\mathcal{H}_{\mathcal{F}_{n+1}}}\right) \mathbb{I}_{\{\mathcal{H}_{\mathcal{F}_n} \leq \mathcal{H}_{\mathcal{F}_{n+1}}\}} \,\bigg|\, \mathcal{F}_n\right] \\
&\stackrel{(a)}{\leq} e\mathbf{E}\left[\frac{\mathcal{H}_{\mathcal{F}_{n+1}}}{\mathcal{H}_{\mathcal{F}_n}} \mathbb{I}_{\{\mathcal{H}_{\mathcal{F}_n} \leq \mathcal{H}_{\mathcal{F}_{n+1}}\}} \,\bigg|\, \mathcal{F}_n\right] \\
&\leq e\mathbf{E}\left[\frac{\mathcal{H}_{\mathcal{F}_{n+1}}}{\mathcal{H}_{\mathcal{F}_n}} \,\bigg|\, \mathcal{F}_n\right] \\
&\stackrel{(b)}{\leq} e
\end{aligned} \qquad (62)$$

where in $(a)$ we have used the fact that $-x \ln x \leq e$ for all $x > 0$ and in $(b)$ we used $\mathbf{E}\left[\mathcal{H}_{\mathcal{F}_{n+1}}|\mathcal{F}_n\right] \leq \mathcal{H}_{\mathcal{F}_n}$. Thus,

$$\begin{aligned}
\mathbf{E}\left[\left|\ln \frac{\mathcal{H}_{\mathcal{F}_n}}{\mathcal{H}_{\mathcal{F}_{n+1}}}\right| \,\bigg|\, \mathcal{F}_n\right] &\leq \mathbf{D}\left(\mathbf{E}\left[\mathcal{S}_{n+1} - \mathcal{S}_n|\mathcal{F}_n\right]\right) + 2e \\
&\leq \mathbf{D}(0) + 2e + \mathbf{D}'(0)\mathbf{E}\left[\mathcal{S}_{n+1} - \mathcal{S}_n|\mathcal{F}_n\right]
\end{aligned} \qquad (63)$$

where $\mathbf{D}'(0) = \frac{d}{d\mathcal{P}}\mathbf{D}(\mathcal{P})\big|_{\mathcal{P}=0}$. Substituting this into the expectation of (59),

$$\mathbf{E}\left[|\ln \mathcal{H}_{\mathcal{F}_{\tau_s}}|\right] \leq \mathbf{D}(0)\mathbf{E}\left[\tau_s\right] + \mathbf{D}'(0)\mathbf{E}\left[\mathcal{S}_{\tau_s}\right] + 2e\mathbf{E}\left[\tau_s\right] + |\ln \ln M|$$

This is finite, verifying the first part of the condition in (57).

Finally we must verify the second part of the condition, i.e., that

$$\lim_{n \to \infty} \mathbf{E}\left[|\mathbf{W}_n|\mathbb{I}_{\{\tau_s \geq n\}}\right] = 0$$

Let $\xi'_n$ be

$$\begin{aligned}
\xi'_n &\triangleq |\mathbf{W}_n|\mathbb{I}_{\{\tau_s \geq n\}} \\
&\leq \left[|\ln \mathcal{H}_{\mathcal{F}_n}| + n\mathbf{D}(\mathcal{P}) + \gamma_\mathbf{D}^\mathcal{P}\left(\mathbf{E}\left[\mathcal{S}_n|\mathcal{F}_n\right] + \mathcal{P}n\right)\right]\mathbb{I}_{\{\tau_s \geq n\}} \\
&\leq \left[\ln \frac{\ln M}{\mathcal{H}_{\mathcal{F}_n}} + |\ln \ln M| + n\mathbf{D}(\mathcal{P}) + \gamma_\mathbf{D}^\mathcal{P}\left(\mathbf{E}\left[\mathcal{S}_n|\mathcal{F}_n\right] + \mathcal{P}n\right)\right]\mathbb{I}_{\{\tau_s \geq n\}} \\
&\leq \left[\sum_{n=0}^{n-1}\left|\ln \frac{\mathcal{H}_{\mathcal{F}_k}}{\mathcal{H}_{\mathcal{F}_{k+1}}}\right| + |\ln \ln M| + n\mathbf{D}(\mathcal{P}) + \gamma_\mathbf{D}^\mathcal{P}\left(\mathbf{E}\left[\mathcal{S}_n|\mathcal{F}_n\right] + \mathcal{P}n\right)\right]\mathbb{I}_{\{\tau_s \geq n\}}
\end{aligned}$$

Following the same set of steps as in (55),

$$\xi'_n \leq \sum_{n=0}^{\tau_s - 1}\left|\ln \frac{\mathcal{H}_{\mathcal{F}_k}}{\mathcal{H}_{\mathcal{F}_{k+1}}}\right| + |\ln \ln M| + \tau_s\mathbf{D}(\mathcal{P}) + \gamma_\mathbf{D}^\mathcal{P}\left(M\mathbf{E}\left[\mathcal{S}_{\tau_s}\right] + \mathcal{P}\tau_s\right) \triangleq \zeta'.$$

Taking the expectation of both sides, using (63), and using the hypotheses $\mathbf{E}[\tau] < \infty$ and $\mathbf{E}[\mathcal{S}_\tau] < \infty$, we see that $\mathbf{E}[\zeta'] < \infty$. Thus, using Lebesgue's dominated convergence theorem, together with the fact that $\lim_{n\to\infty} \mathbf{P}[\xi'_n = 0] = 1$ implies that $\lim_{n\to\infty} \mathbf{E}[\xi'_n] = 0$. **QED**

**Proof of Lemma 11:**

We use the same shorthand notation as in the proof of Lemma 7 to upper bound $\ln \mathcal{H}_{\mathfrak{f}_n} - \ln \mathcal{H}_{\mathfrak{f}_{n+1}}$. Let $Y_{n+1} = j$.

$$\ln \mathcal{H}_{\mathfrak{f}_n} = \ln \sum_{i=1}^{M} p(i) \ln \frac{1}{p(i)} \qquad \ln \mathcal{H}_{\mathfrak{f}_{n+1}} = \ln \sum_{i=1}^{M} p(i|j) \ln \frac{1}{p(i|j)}$$

Note that

$$\frac{p(i)}{p(i|j)} = \frac{\psi(j)}{\psi(j|i)} \leq \max_{k,m} \frac{P_{kj}}{P_{mj}} \tag{64}$$

where we have used the fact that both $\psi(j)$ and $\psi(j|i)$ are in the convex hull of the set of transition probabilities $P_{kj}$. Using the non-negativity of the divergence followed by (64),

$$\sum_i p(i) \ln \frac{1}{p(i)} \leq \sum_i p(i) \ln \frac{1}{p(i|k)} \leq \left( \max_{k,m} \frac{P_{kj}}{P_{mj}} \right) \sum_i p(i|j) \ln \frac{1}{p(i|j)}$$

Including $j$ in the maximization above, this is valid for all possible outputs $Y_{n+1}$ and is thus equivalent to (35). **QED**

**Proof of the Converse, Theorem 4:**

Theorem 4 will be proved for the discrete-time memoryless channels defined in Section 4. This includes DMC's with $P_{kj} > 0$ for all $k, j$ as a special case. The discussion in Subsection 3.4.1 is valid except for $\ln \mathcal{H}_{\mathcal{F}_{\tau_1}} \geq -\mathbf{F}$ and the consequent inequality (36). As a substitute, we will use Assumption 2 of Section 4 to show that $\mathbf{E}[\tau - \tau_1]$ can be lower bounded, for each $\Delta > 0$, by

$$\forall \Delta > 0 \; \mathbf{E}[\tau - \tau_1] \geq \frac{-\ln P_e - \ln[\ln M - \ln P_e + 1] - \Delta - \xi(\Delta)(1+\mathcal{P})\mathbf{E}[\tau]}{\mathbf{D}(\mathcal{P}_2)} \tag{65}$$

Proof of equation 65 will be presented subsequent to the current proof.

Assume that the theorem is false. Then a sequence of codes, indexed by superscript $i$, exists such that the durations $\tau^i$ satisfy $\lim_{i\to\infty} \mathbf{E}[\tau^i] = \infty$ and each code satisfies all the conditions above but violates (37). Define $\eta^i = \frac{\mathbf{E}[\tau_1^i]}{\mathbf{E}[\tau^i]}$. Then using (32) for the $i$th code and dividing both sides by $\mathbf{E}[\tau^i]$, we get

$$\eta^i \geq \frac{(\ln M^i)\left[1 - P_e^i(\ln M^i - \ln P_e^i + 1) - \frac{1}{\ln M^i}\right]}{\mathbf{E}[\tau^i]\,\mathbf{C}(\mathcal{P}_1^i)}$$

$$\geq \frac{(R+\delta)\left[1 - P_e^i(\ln M^i - \ln P_e^i + 1) - \frac{1}{\ln M^i}\right]}{\mathbf{C}(\mathcal{P}_1^i)}$$

The term in brackets above approaches 1 since $P_e^i \leq \exp\{-\mathbf{E}[\tau^i][E(R,\mathcal{P}) + \delta]\}$ and $\ln(M^i)$ lies between $\mathbf{E}[\tau^i](R+\delta)$ and $\mathbf{E}[\tau^i]\mathbf{C}(\mathcal{P})$. Thus

$$\eta^i \geq \frac{R}{\mathbf{C}(\mathcal{P}_1^i)} \qquad \text{for sufficiently large } i, \tag{66}$$

Similarly, dividing both sides of (65) by $\mathbf{E}\left[\tau^i\right]$,

$$
\begin{aligned}
1 - \eta^i &\geq \frac{-\ln(P_e^i) - \ln[\ln M - \ln P_e^i + 1] - \Delta - (1+\mathcal{P})\xi(\Delta)\mathbf{E}\left[\tau\right]}{\mathbf{E}\left[\tau\right]\mathbf{D}(\mathcal{P}_2^i)} \\
&\geq \frac{(E(R,\mathcal{P})+\delta)\left[1 - \frac{\ln[\ln M - \ln P_e^i + 1]}{\ln(1/P_e^i)}\right] - \frac{\Delta}{\mathbf{E}[\tau^i]} - (1+\mathcal{P})\xi(\Delta)}{\mathbf{D}(\mathcal{P}_2^i)}
\end{aligned}
$$

Let $\Delta^*$ be such that $(1+\mathcal{P})\xi(\Delta^*) \leq \delta/4$. Then for sufficiently large

$$1 - \eta^i \geq \frac{E(R,\mathcal{P}) + \delta/2}{\mathbf{D}(\mathcal{P}_2^i)} \tag{67}$$

From (66), $\mathcal{P}_1^i \geq \mathbf{C}^{-1}(R/\eta^i)$, so from the energy constraint $\mathbf{E}\left[\mathcal{S}_\tau\right] \leq \mathcal{P}\mathbf{E}\left[\tau\right] + \delta$, we have

$$\mathcal{P}_2^i \leq \frac{\mathcal{P} - \mathbf{C}^{-1}(\frac{R}{\eta^i}) + \frac{\delta}{\mathbf{E}[\tau^i]}}{1 - \eta^i} \tag{68}$$

Thus, using (67) and (68) for large enough $\overline{\tau}^i$,

$$
\begin{aligned}
E(R,P) + \delta/2 &\leq (1-\eta^i)\mathbf{D}\left(\frac{\mathcal{P} - \eta^i \mathbf{C}^{-1}(\frac{R}{\eta^i})}{1-\eta^i}\right) + \delta/4 \\
&\leq E(R,P) + \delta/4, \tag{69}
\end{aligned}
$$

where the definition of $E(R,P)$ in (12) is used to establish a contradiction. **QED**

**Proof of Inequality 65:**

For any channel and code, let $v_n$ be the random variable $v_n = \ln \mathcal{H}_{\mathcal{F}_n} - \ln \mathcal{H}_{\mathcal{F}_{n+1}}$. Assumption 2 of section 4 asserts that there is a function $\xi(\Delta)$ satisfying $\lim_{\Delta \to \infty} \xi(\Delta) = 0$ such that for all $n$ and $\Delta \geq 0$,

$$\mathbf{E}\left[v_n \mathbb{I}_{\{v_n \geq \Delta\}} \mid \mathcal{F}_n\right] \leq \xi(\Delta)(1 + \mathbf{E}\left[\mathcal{S}_{n+1} - \mathcal{S}_n \mid \mathcal{F}_n\right]) \tag{70}$$

For all sample values $\mathfrak{f}_n$ of $\mathcal{F}_n$ such that $\mathcal{H}_{\mathfrak{f}_n} > 1$, we see that $v_n \geq -\ln \mathcal{H}_{\mathcal{F}_{n+1}}$. It follows from this that $\mathbb{I}_{\{v_n \geq \Delta\}} \geq \mathbb{I}_{\{-\ln \mathcal{H}_{\mathcal{F}_{n+1}} \geq \Delta\}}$ and thus that

$$v_n \mathbb{I}_{\{v_n \geq \Delta\}} \geq -\ln \mathcal{H}_{\mathcal{F}_{n+1}} \mathbb{I}_{\{-\ln \mathcal{H}_{\mathcal{F}_{n+1}} \geq \Delta\}}$$

Substituting this into (70) for all such $\mathfrak{f}_n$,

$$\mathbf{E}\left[-\ln \mathcal{H}_{\mathcal{F}_{n+1}} \mathbb{I}_{\{-\ln \mathcal{H}_{\mathcal{F}_{n+1}} \geq \Delta\}} \middle| \mathcal{F}_n = \mathfrak{f}_n\right] \leq \xi(\Delta)(1 + \mathbf{E}\left[\mathcal{S}_{n+1} - \mathcal{S}_n \mid \mathcal{F}_n = \mathfrak{f}_n\right]) \tag{71}$$

Note that $\mathcal{H}_{\mathcal{F}_n} > 1$ holds, and thus (71) also holds, for each $\mathfrak{f}_n$ such that $n < \tau_1$. Thus, (71) will hold for all $\mathfrak{f}_n$ if the indicator function $\mathbb{I}_{\{n < \tau_1\}}$ is inserted on both sides, i.e.,

$$
\begin{aligned}
\mathbf{E}\left[-\leq \mathcal{H}_{\mathcal{F}_{n+1}} \mathbb{I}_{\{-\ln \mathcal{H}_{\mathcal{F}_{n+1}} \geq \Delta\}} \mathbb{I}_{\{n<\tau_1\}} \middle| \mathcal{F}_n\right] &\leq \xi(\Delta)\mathbf{E}\left[(1 + \mathcal{S}_{n+1} - \mathcal{S}_n)\mathbb{I}_{\{n<\tau_1\}} \middle| \mathcal{F}_n\right]) \\
\mathbf{E}\left[-\ln \mathcal{H}_{\mathcal{F}_{n+1}} \mathbb{I}_{\{-\ln \mathcal{H}_{\mathcal{F}_{n+1}} \geq \Delta\}} \mathbb{I}_{\{n<\tau_1\}}\right] &\leq \xi(\Delta)\mathbf{E}\left[(1 + \mathcal{S}_{n+1} - \mathcal{S}_n)\mathbb{I}_{\{n<\tau_1\}}\right]) \tag{72}
\end{aligned}
$$

where we have taken the expected value over $\mathcal{F}_n$ on both sides. Note that

$$\mathbb{I}_{\{-\ln \mathcal{H}_{\mathcal{F}_{n+1}} \geq \Delta\}} \mathbb{I}_{\{n < \tau_1\}} = \mathbb{I}_{\{-\ln \mathcal{H}_{\mathcal{F}_{n+1}} \geq \Delta\}} \mathbb{I}_{\{n+1 = \tau_1\}}$$

For any $k > 0$, we next sum (72) over $0 \leq n < k$.

$$\mathbf{E}\left[-\ln \mathcal{H}_{\mathcal{F}_{\min(k,\tau_1)}} \mathbb{I}_{\{-\ln \mathcal{H}_{\mathcal{F}_{\min(k,\tau_1)}} \geq \Delta\}}\right] \leq \xi(\Delta) \mathbf{E}\left[\min(k,\tau_1) + \mathcal{S}_{\min(k,\tau_1)}\right]$$

Using

$$|\ln \mathcal{H}_{\mathcal{F}_{\tau_1,l}}| \mathbb{I}_{\{\mathcal{H}_{\mathcal{F}_{\tau_1,l}} \leq \exp(-\Delta)\}} \leq |\ln \mathcal{H}_{\mathcal{F}_{\tau_1}}| \mathbb{I}_{\{\mathcal{H}_{\mathcal{F}_{\tau_1}} \leq \exp(-\Delta)\}}$$

together with $\mathbf{E}\left[|\ln \mathcal{H}_{\mathcal{F}_{\tau_1}}|\right] < \infty$, $\mathbf{E}[\tau_1] < \infty$ and Lebesgue's dominated convergence theorem, one can show that,

$$\mathbf{E}\left[\ln \mathcal{H}_{\mathcal{F}_{\tau_1}} \mathbb{I}_{\{\mathcal{H}_{\mathcal{F}_{\tau_1}} \leq \exp(-\Delta)\}}\right] \geq -\xi(\Delta)(\mathbf{E}[\tau_1] + \mathbf{E}[\mathcal{S}_{\tau_1}]) \tag{73}$$

Consequently

$$\mathbf{E}\left[\ln \mathcal{H}_{\mathcal{F}_{\tau_1}}\right] \geq -\Delta - \xi(\Delta)(\mathbf{E}[\tau_1] + \mathbf{E}[\mathcal{S}_{\tau_1}]) \tag{74}$$

Now recall the inequality (34)

$$\mathbf{E}[\ln \mathcal{H}_{\mathcal{F}_\tau}] \leq \ln[P_e(\ln M - \ln P_e + 1)]$$

Using the Lemma 10, we get

$$\begin{aligned}
\mathbf{E}[\tau - \tau_1] &\geq \mathbf{E}\left[\ln \mathcal{H}_{\mathcal{F}_{\tau_1}} - \ln \mathcal{H}_{\mathcal{F}_\tau}\right] \mathbf{D}(\mathcal{P}_2) \\
&\geq \frac{-\ln P_e - \ln[\ln M - \ln P_e + 1] - \Delta - \xi(\Delta)(\mathbf{E}[\tau_1] + \mathbf{E}[\mathcal{S}_{\tau_1}])}{\mathbf{D}(\mathcal{P}_2)} \\
&\geq \frac{-\ln P_e - \ln[\ln M - \ln P_e + 1] - \Delta - \xi(\Delta)(1 + \mathcal{P}) \mathbf{E}[\tau]}{\mathbf{D}(\mathcal{P}_2)}
\end{aligned}$$

**QED**

# References

[1] B. Nakiboḡlu, R.G. Gallager, and M. Win. Error exponents for variable-length block codes with feedback and cost constraints. In *ISIT 2006, Seattle, Washington July 9- 14, 2006*, 2006.

[2] M. V. Burnashev. Data transmission over a discrete channel with feedback, random transmission time. *Problemy Perdachi Informatsii*, 12, No. 4:10–30, 1976.

[3] M. V. Burnashev. Sequential discrimination of hypotheses with control of observations. *Math. USSR Izvestija*, 15, No. 3:419–440, 1980.

[4] E.R. Berlekamp C.E. Shannon, R.G. Gallager. Lower bounds to error probability for coding on discrete memoryless channels. *Information and Control*, 10, No. 1:65–103, 1967.

[5] R. L. Dobrushin. An asymptotic bound for the probability error of information transmission through a channel without memory using the feedback. *Problemy Kibernetiki*, vol 8:161–168, 1962.

[6] P. Elias. Channel capaccity without coding. Quarterly report, MIT Research Lab Of Electronics, Research Lab Of Electronics, Cambridge, Massachusetts, October 1956.

[7] Robert G. Gallager. *Information Theory and Reliable Communication*. John Wiley & Sons, Inc., New York, NY, USA, 1968.

[8] E. A. Haroutunian. A lower bound of the probability of error for channels with feedback. *Problemy Peredachi Informatsii*, vol 13:36–44, 1977.

[9] A. Kramer. Improving communication reliability by use of an intermittent feedback channel. *Information Theory, IEEE Transactions on*, Vol.15, Iss.1:52–60, 1969.

[10] M. S. Pinsker. The probability of error in block transmission in a memoryless gaussian channel with feedback. *Problemy Perdachi Informatsii*, 4(4):1–4, 1968.

[11] A. Sahai. Why block length and delay are not the same thing. preprint.

[12] A. Sahai and T. Şimşek. On the variable-delay reliability function of discrete memoryless channels with access to noisy feedback. In *ITW2004, San Antonio, Texas, October 24 29, 2004*, 2004.

[13] J. P. M. Schalkwijk. A coding scheme for additive noise channels with feedback–ii: Bandlimited signals. *Information Theory, IEEE Transactions on*, Vol.12, Iss.2:183–189, 1966.

[14] J. P. M. Schalkwijk and T. Kailath. A coding scheme for additive noise channels with feedback–i: No bandwidth constraint. *Information Theory, IEEE Transactions on*, Vol.12, Iss.2:172–182, 1966.

[15] S.Chang. Theory of information feedback systems. *IEEE Transactions on Information Theory*, Vol 2, Iss 3:29–40, 1956.

[16] C. Shannon. The zero error capacity of a noisy channel. *Information Theory, IEEE Transactions on*, Vol. 2, Iss 3:8–19, 1956.

[17] Albert N. Shiriaev. *Probability*. Springer-Verlag Inc., New York, NY, USA, 1996.

[18] H. Yamamoto and K. Itoh. Asymptotic performance of a modified schalkwijk-barron scheme for channels with noiseless feedback. *IEEE Transactions on Information Theory*, Vol.25, Iss.6:729–733, 1979.